\documentclass[a4paper,fleqn,usenatbib]{mnras}            % for single column format 
\usepackage{graphicx}
\usepackage{caption}
\usepackage{subcaption}
\usepackage{float}
\usepackage{epstopdf}
\usepackage{natbib}
\usepackage{url}
\usepackage{etoolbox}
\makeatletter
\patchcmd\@combinedblfloats{\box\@outputbox}{\unvbox\@outputbox}{}{%
   \errmessage{\noexpand\@combinedblfloats could not be patched}%
}%
 \makeatother
\title[UVIT-HST-GAIA view of NGC 288]{UVIT-HST-GAIA  view of NGC 288: A census of hot stellar population and their properties from UV}

\author[Sahu et al.]{
Snehalata Sahu,$^{1}$\thanks{E-mail: snehalata@iiap.res.in}
Annapurni Subramaniam,$^{1}$
Patrick C\^ot\'e,$^{2}$
N. Kameswara Rao,$^{1}$ 
\newauthor and Peter B. Stetson$^{2}$\\
$^{1}$Indian Institute of Astrophysics, Bengaluru, 560034, India\\
$^{2}$National Research Council of Canada, Herzberg Astronomy and Astrophysics Research Centre, Victoria, Canada\\
}

\date{Accepted XXXX. Received YYYY; in original form ZZZZ}

\pubyear{2018}

\begin{document}
\label{firstpage}
\pagerange{\pageref{firstpage}--\pageref{lastpage}}
\maketitle
%
% Please do not change the following 6 lines, sizes stolen 
% from XMM proposal science justification.
\textheight=247mm
\textwidth=180mm
\topmargin=-7mm
\oddsidemargin=-10mm
\evensidemargin=-10mm
\parindent 14pt
%

%
%---- ENTRY 1 ----------------------------------------------------------
%
% Type below, within the curly braces{}, the title of your proposal
% Abstract of the paper
\begin{abstract}
A complete census of Blue Horizontal Branch (BHB) and Blue Straggler Star (BSS) population within the 10$\arcmin$ radius from the center of the Globular Cluster, NGC 288 is presented, based on the images from the Ultraviolet Imaging Telescope (UVIT). The UV and UV$-$optical Colour-Magnitude Diagrams (CMDs) are constructed by combining the UVIT, HST-ACS and ground data and compared with the BaSTI isochrones generated for UVIT filters. We used stellar proper motions data from GAIA DR2 to select the cluster members. Our estimations of the temperature distribution of 110 BHB stars reveal two peaks with the main peak at $T_{eff}\sim$ 10,300 K with the distribution extending up to $T_{eff}\sim$ 18,000 K. We identify the well known photometric gaps including the G-jump in the BHB distribution which are located between the peaks. We detect a plateau in the FUV magnitude for stars hotter than $T_{eff}\sim$ 11,500 K (G-jump), which could be due to the effect of atomic diffusion. We detect 2 Extreme HB (EHB) candidates with temperatures ranging from 29,000 to 32,000 K. The radial distribution of 68 BSSs suggests that the bright BSSs are more centrally concentrated than the faint BSS and the BHB distribution. We find that the BSSs have a mass range of 0.86 - 1.25 M$_{\odot}$ and an age range of 2 - 10 Gyr with a peak at 1 M$_{\odot}$ and 4 Gyr respectively. This study showcases the importance of combining UVIT with HST, ground, and GAIA data in deriving HB and BSS properties.
\end{abstract}


\section{Introduction}
Globular clusters (GCs) are among the oldest systems in our galaxy with central densities varying from 10 to 10$^{6}$ $L_{\odot}/pc^{3}$ \citep{harris1996, harris2010}. They provide the best platforms to study the properties of exotic stellar populations such as Blue Straggler Stars (BSSs), close binary systems, Cataclysmic Variables (CVs), low mass X-ray binaries (LMXBs) resulting from the dynamical interactions such as collisions or mergers between the members of the cluster. NGC 288 is a low density GC ($\rho_{c}\sim$ 60.25 $L_{\odot}/pc^{3}$, \cite{mc2005}) located in the Constellation Sculptor. It is of intermediate metallicity with [Fe/H] = $-$1.3 \citep{Carretta2009}. NGC 288 is known to be located close to the South Galactic Pole, with a retrograde orbit \citep{Dinescu1997} and an evolution strongly driven by the galactic tidal field \citep{leon2000}. 

Historically, NGC 288 is considered as a peculiar GC, with the presence of a purely blue Horizontal Branch (HB) stars, coupled with a relatively high metallicity \citep{Cannon1974, Buon1984}. The peculiarities found in the HB morphology are well known, one of which being the second parameter problem \citep{sandage1960, sandage1967, van1993}, which refers to the observation that parameters other than metallicity, such as age and/or He abundance, affects the colour distribution of HB stars. NGC 288 and NGC 362 form one of the best known $``$second-parameter pair$"$ of GCs \citep{catelan2001, Bella2001}, where NGC 362 presenting a very red HB, a complete opposite of the blue HB of NGC 288, given that these clusters have similar chemical compositions. \cite{catelan2001} found that when the overall HB morphology of the two clusters can be reproduced with an age difference of 2 Gyr, the details are not fitted easily. 

In clusters with a relatively broad HB distribution, a number of discontinuities or jumps have been identified, though this has a dependence on the band passes used \citep{Brown2016}. Some of the well known jumps are $``$Grundahl jump$"$ (G-jump) within the blue HB (BHB) at $\sim$ 11,500 K \citep{Grundahl1999} and the $``$Momany Jump$"$ (M-jump) within the extreme HB (EHB) at $\sim$ 23,000 K \citep{Momany2002, Momany2004}, and these were detected in many GCs with sufficient number of BHB and EHB stars \citep{Ferraro1998}. \cite{Brown2016} demonstrated that the HB discontinuities are remarkably consistent in temperature, with the help of blue and UV HST photometry. The atmospheric processes are found to play a major role in the making of the HB discontinuities. The BHB stars hotter than the G-jump exhibit metal abundances enhanced via radiative levitation and He abundances diminished via gravitational settling \citep{Moehler1999, Moehler2000, Behr2003, Pace2006}. \cite{Khalack2010} found observational evidence for vertical stratification of iron, which supports the efficiency of atomic diffusion among the hotter BHB stars, which include three BHB stars in NGC 288. \cite{Moehler2014} also found evidence for the presence of diffusion among the BHB stars hotter than $T_{eff}\sim$ 11,500 K.

It is well known that the study of binary populations in globular clusters can provide powerful constraints on both dynamical models and models of formation of exotic objects such as BSSs. BSSs are hydrogen burning stars located above the MS in the optical Colour-Magnitude Diagrams (CMDs). BSSs were first discovered by \cite{Sandage} in the optical CMD of GC M3. Two formation mechanisms have been proposed by the previous studies- stellar collisions leading to mergers \citep{Hills, Chatterjee2013} and mass transfer in the binary systems \citep{Mccrea, Chen2008, Knigge2009, Leigh2013}. The first mechanism is most favoured in dense cluster environments such as central regions of the cluster whereas the second mechanism is preferred in low density environments. \cite{Bellazzini2000} discovered a binary sequence in the optical CMD of NGC 288 from the HST-WFPC2 observations of the cluster. Later,  \cite{Bellazzini2001} studied the binary systems and BSSs in the cluster with HST-WFPC2 survey data and estimated the binary fraction ($f_{b}$) to range from 8$\%$ to 38$\%$. They found that most of the binary systems are located within the half-light radius of the cluster. They also found a high specific frequency of BSSs where they used ($m_{F255W}-m_{F336W}$) vs $m_{F255W}$ plane for selecting the BSS population of the cluster. They concluded that the mass transfer between the primordial binary systems might have led to the formation of BSSs which is as efficient in the low density environments as the collision mechanisms in the high density environments of the GCs. As the BSS population are more massive than the majority of the stars in the cluster, they are affected by the dynamical friction. \cite{Ferraro2012} demonstrated that the morphology of the normalised radial distribution of the BSS population is strongly shaped by the action of dynamical friction. The normalised BSS distribution can thus be used as a $``$dynamical clock$"$, which reflects the dynamical evolutionary stage of the cluster, in comparison to its stellar evolutionary age.  

Recent studies of GCs \citep{Ferraro2003, Haurberg2010, Dieball2010, Schiavon2012, Piotto2015, Parada2016, Raso2017, Dieball2017}  have shown that Ultra-Violet (UV) CMDs are important tools for identifying and studying the properties of UV bright stellar populations such as BSSs and HBs. \cite{Schiavon2012} generated FUV$-$NUV vs FUV CMD of 44 GCs using GALEX data. They found that the HBs form a diagonal sequence in the UV CMD with BSSs running in a parallel sequence lying just 1 mag below the HBs. They also identified and catalogued the UV bright candidates such as post-AGB (PAGB) stars thus, highlighting the importance of UV observations in GCs.

The HST Far UV observations of 3 GCs by \cite{Ferraro1998}  showed that the FUV CMDs are very helpful in identifying the HB peculiarities such as gaps which can throw light on the physical processes that HB stars undergo during the mass loss in the Red Giant Branch (RGB) phase. Many studies such as \cite{Dalessandro2011, Lagioia2015, Brown2016}, showed that the FUV CMDs are very sensitive to the temperature variations in the HBs and thus, are best to study the HB morphology and the temperatures of HB stars as compared to the optical CMDs. %   UV photometry is also important in the study of EHB and Blue Hook stars as optical colours become insensitive at the temperatures of EHB stars. 

UVIT study of the GC NGC 1851 by \cite{Subramaniam2017} showed UVIT's capability in identifying the multiple stellar populations by analysing the HB morphology in the FUV and NUV CMDs of the cluster. Another UVIT study of NGC 188 by \cite{Subramaniam1-2016} detected the presence of a PAGB companion to one of the BSS with the help of the SED generated by combining UV, optical and IR fluxes of the BSS. Thus, owing to its excellent spatial resolution ($\sim1\farcs2$) and large field of view (FOV $\sim 28'$), UVIT observations are very useful for extracting the properties of UV bright stellar populations in a star cluster.

In this work, we present the results of a UVIT imaging study of the GC NGC 288 using three filters (F148W, F169M and N279N) of UVIT. We compare our data with HST-Advanced Camera Survey (ACS) data \citep{Sarajedini2007} and ground data. We used the proper motions data available from GAIA Data Release 2 (DR2) \citep{Gaia2018b} to select the cluster members detected by UVIT.

The paper is organised as follows. We describe the observations and data reductions in Section \ref{obs}. In Section \ref{uvocmd}, we present the UV and optical CMDs with the discussions on various stellar populations and their membership from GAIA. Sections \ref{temp_bhb} and  \ref{temp_ehb} describe the temperature distribution of BHB and EHB stars with Section \ref{param_bss} focusing on the properties of BSS derived from the UVIT photometry along with the subsequent discussions in each sections. We summarise and conclude our results in Section \ref{conclude}.

\begin{table*}
\centering
\caption{Observation and Photometry details of NGC 288}
\resizebox{125mm}{!}{

\begin{tabular}{cccccc}
\hline
\hline
Filter & $\lambda_{mean}$ & $\Delta \lambda$ &  Exposure Time & Zero point & Number of Sources \\
& [$\mathrm{\AA}$] & [$\mathrm{\AA}$] & [sec] & [mag] &\\\hline
F148W  & 1481 & 500 & 7057 & 18.00  & 258\\
F169M  & 1608 & 290 & 4573 & 17.45 & 293 \\
N279N  & 2792 & 90 & 14778 & 16.46 & 5141\\\hline
\label{phot}
\end{tabular}
}
\end{table*}

\begin{figure*}
\begin{center}
\includegraphics[scale=0.45]{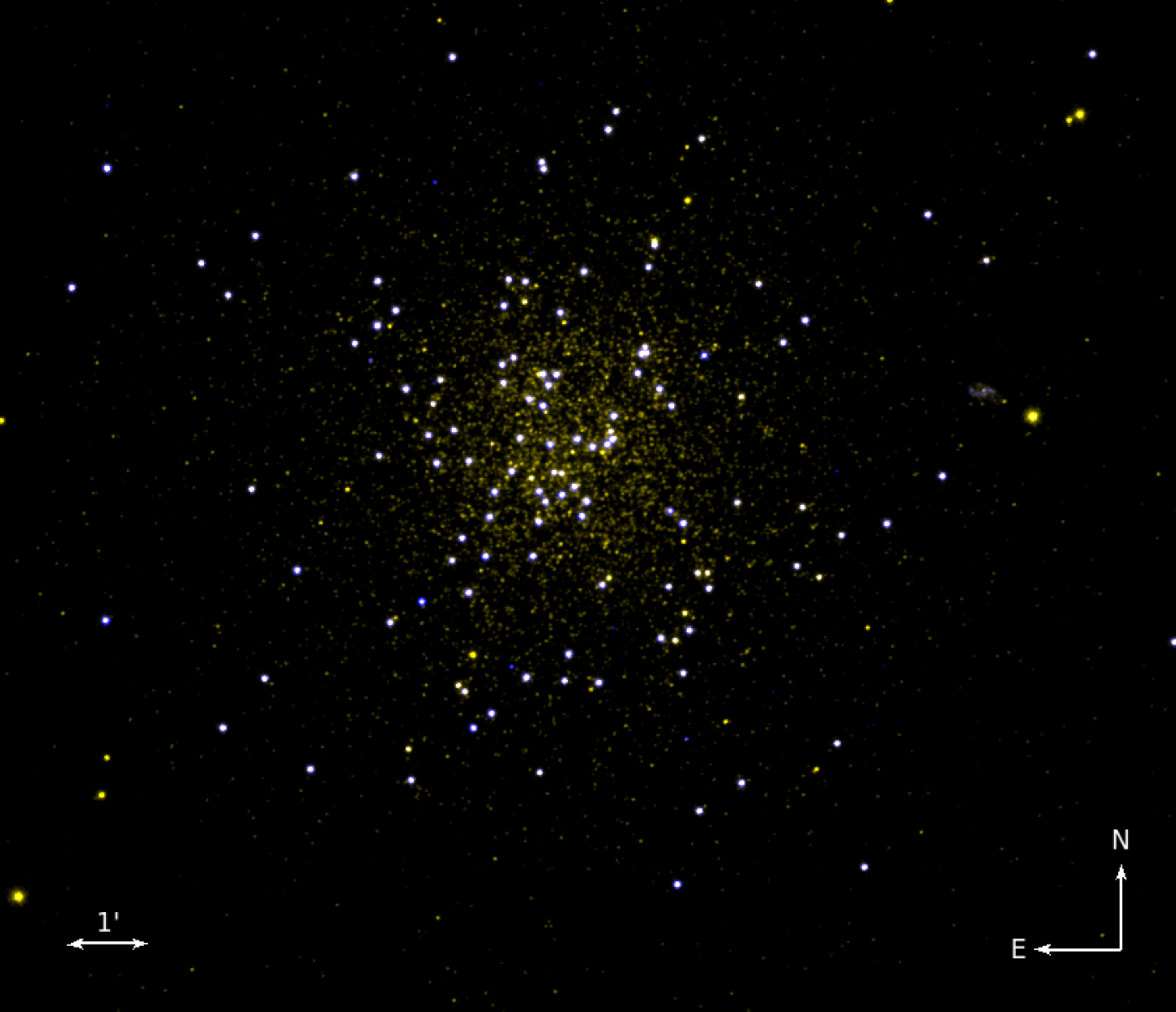}
\caption{UVIT image of NGC 288 where blue corresponds to the FUV detections and yellow corresponds to the NUV detections acquired with F148W and N279N filters of UVIT respectively.}
\label{uvit_image}
\end{center}
\end{figure*}

\section{Observations and Data Reductions}\label{obs}

The data presented in this paper are obtained from UVIT instrument on-board the Indian space observatory, ASTROSAT. The UVIT instrument consists of two 38-cm telescopes - one for the FUV and other for the NUV and visible bands. It has a circular field of view $\sim$ 28$'$ in diameter. It collects data in three channels simultaneously, in FUV, NUV and Visible bands  corresponding to $\lambda$ = 1300 - 1800 $\mathrm{\AA}$, 2000 - 3000 $\mathrm{\AA}$ and 3200 - 5500 $\AA$ respectively. The visible channel is mainly used for drift correction. UV detectors work in photon counting mode whereas the visible detector in integration mode. Each band is further subdivided into multiple filters. The effective area curves of the filters are available in the UVIT website\footnote{\url{http://uvit.iiap.res.in/Instrument/Filters}}. Full details of the instrument and calibration results can be found in \cite{Subramaniam2016, Tandon2017}.

\begin{figure*}
\begin{center}
\includegraphics[scale=0.36]{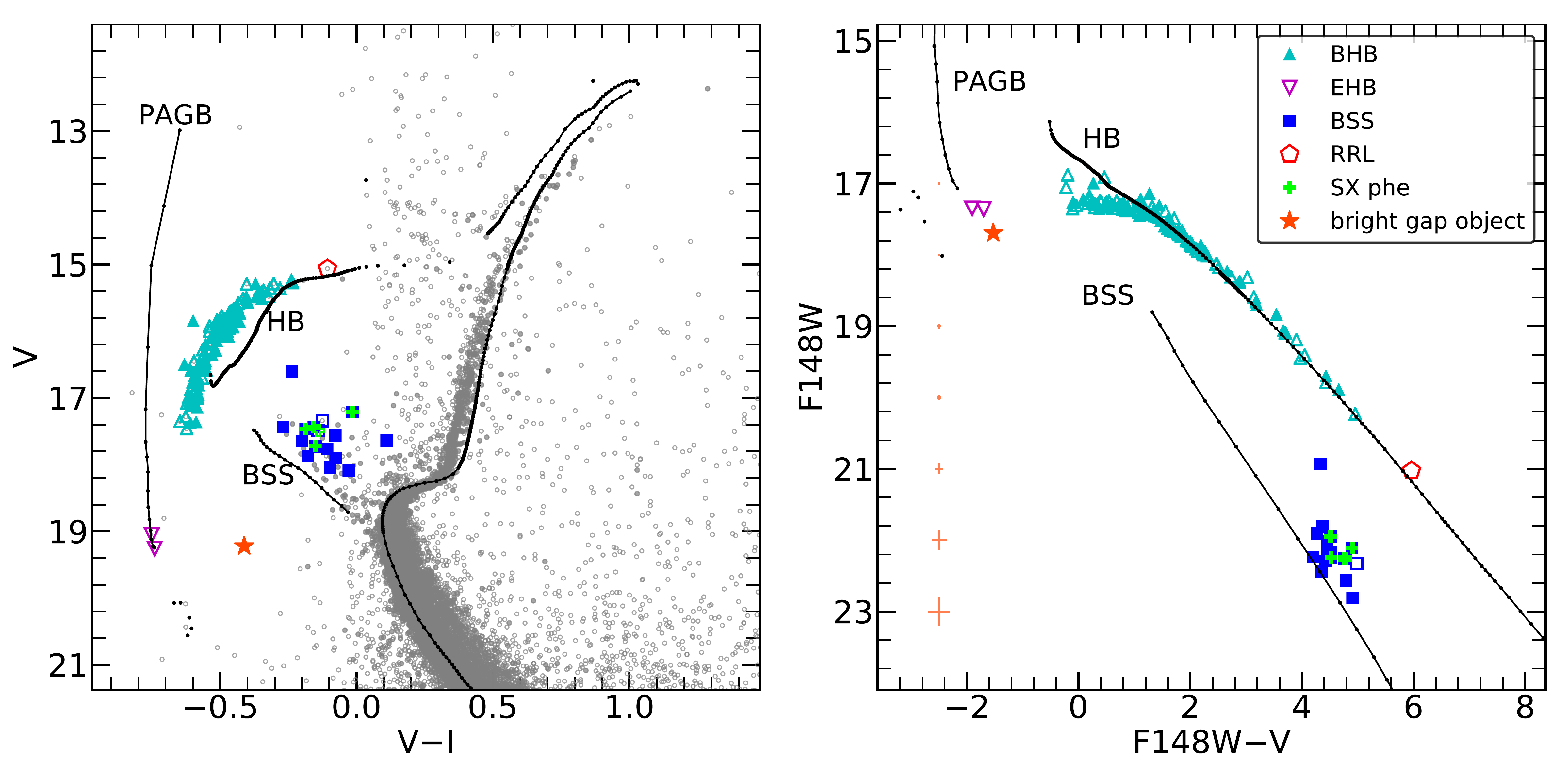}\\
\caption{F148W$-$V vs F148W CMD (right panel) CMD along with the corresponding optical CMD (left panel). The photometric errors in the colour at different magnitude are shown as light red horizontal lines in the right panel. The filled symbols are UVIT cross-matched HST detections and open symbols are UVIT cross-matched ground detections in both the panels. The observed magnitudes and colours are corrected for extinction and reddening. The Figures are over plotted with a BaSTI isochrone (black line and dots) of age 12.6 Gyr \citep{Kaiser2017}, [Fe/H] = $-$1.28 \citep{Carretta2009} and distance modulus = 14.84 \citep{Bella2001}. The various stellar populations marked in the legend are BHB - Blue Horizontal Branch (cyan triangle), EHB - Extreme Horizontal Branch (magenta triangle), BSS - Blue Straggler Star (blue squares), RRL - RR Lyrae (red pentagon) and SX Phe - SX Phoenicis (green plus) variables.}
\label{caf2_v_cmd}
\end{center}
\end{figure*}

\subsection{Photometry}
The cluster was observed during 20-21 August 2016 as a part of the Guaranteed Time (GT) proposal (G05\_009). The images were acquired in three filters of UVIT - F148W and F169M (FUV channel) and N279N (NUV channel). A customised software package, CCDLAB \citep{Postma2017} was used to correct for the spacecraft drift, geometric distortion and flat field correction in the images. The false colour UVIT image of the cluster is shown in Figure \ref{uvit_image}. The observation and photometry details of NGC 288 UVIT images are given in Table \ref{phot}.

We have performed crowded field photometry on the UVIT images using DAOPHOT software package of IRAF/NOAO \citep{Stetson1987} and obtained the magnitudes of sources at several apertures up to four times the FWHM using DAOPHOT task {\it phot}. These magnitudes are based on simple aperture photometry which usually fails to give correct magnitudes for sources in the crowded field. Therefore, we have performed point-spread function (PSF) photometry, where the PSF, created by choosing a few isolated stars in the field is used in the ALLSTAR task to obtain the PSF fitted magnitudes. The FWHM of the PSF are found to be $\sim1\farcs5$, $1\farcs6$ and $1\farcs2$ in F148W, F169M and N279N filters respectively. We have estimated the aperture correction value in each filter using Curve of Growth analysis technique and applied it on PSF generated magnitudes. Finally, we have done saturation correction \citep{Tandon2017} to obtain the final magnitudes in each filter.

We have adopted a reddening value E(B$-$V)=0.03 mag \citep{Ferr1999} to correct the observed magnitudes and colours in all the bands. By considering Fitzpatrick extinction law \citep{Fitzpatrick1999}, we have calculated the extinction coefficients which are found to be R(F148W) = 8.40, R(F169M) = 7.91 and R(N279N) = 5.96 in F148W, F169M and N279N filters of UVIT respectively. For the rest of the paper, the considered magnitudes (AB system) are corrected for extinction and the colours for reddening.  

\section{UVIT and Optical Colour-Magnitude Diagrams}\label{uvocmd}
In order to identify the stellar populations detected with UVIT in different wavebands, we have used the data from the HST-ACS Survey of GCs \citep{Sarajedini2007} and ground (Peter Stetson in private comm.). The HST-ACS pointing covers a region $\sim3\farcm4 \times 3\farcm4$ and the catalogue contains V and I magnitudes obtained from the observations in F606W and F814W filters. We have used Vg and Ig magnitudes given in the catalogue which are calibrated to ground photometry \citep{Sirianni2005}. The advantage of HST-ACS data is that, it has resolved the cluster centre in optical but its FOV covers only the cluster central regions. Therefore, we have considered only sources in the ground data which lie in a region outside the FOV of HST-ACS. The ground data contains U, B, V, R and I magnitudes. We have merged both the HST-ACS data and the ground data in such a way that it covers the full cluster region ($\sim$ 10$'$ in radius from the cluster centre). As the UVIT images have resolved the centre of the cluster, we have detected all the HB and BSS stars in the cluster within 10$'$ radius, in the FUV and NUV filters. Thus, this study presents the identification and analysis of the complete HB and BSS sample in this cluster.

\begin{figure}
\begin{center}
\includegraphics[scale=0.34]{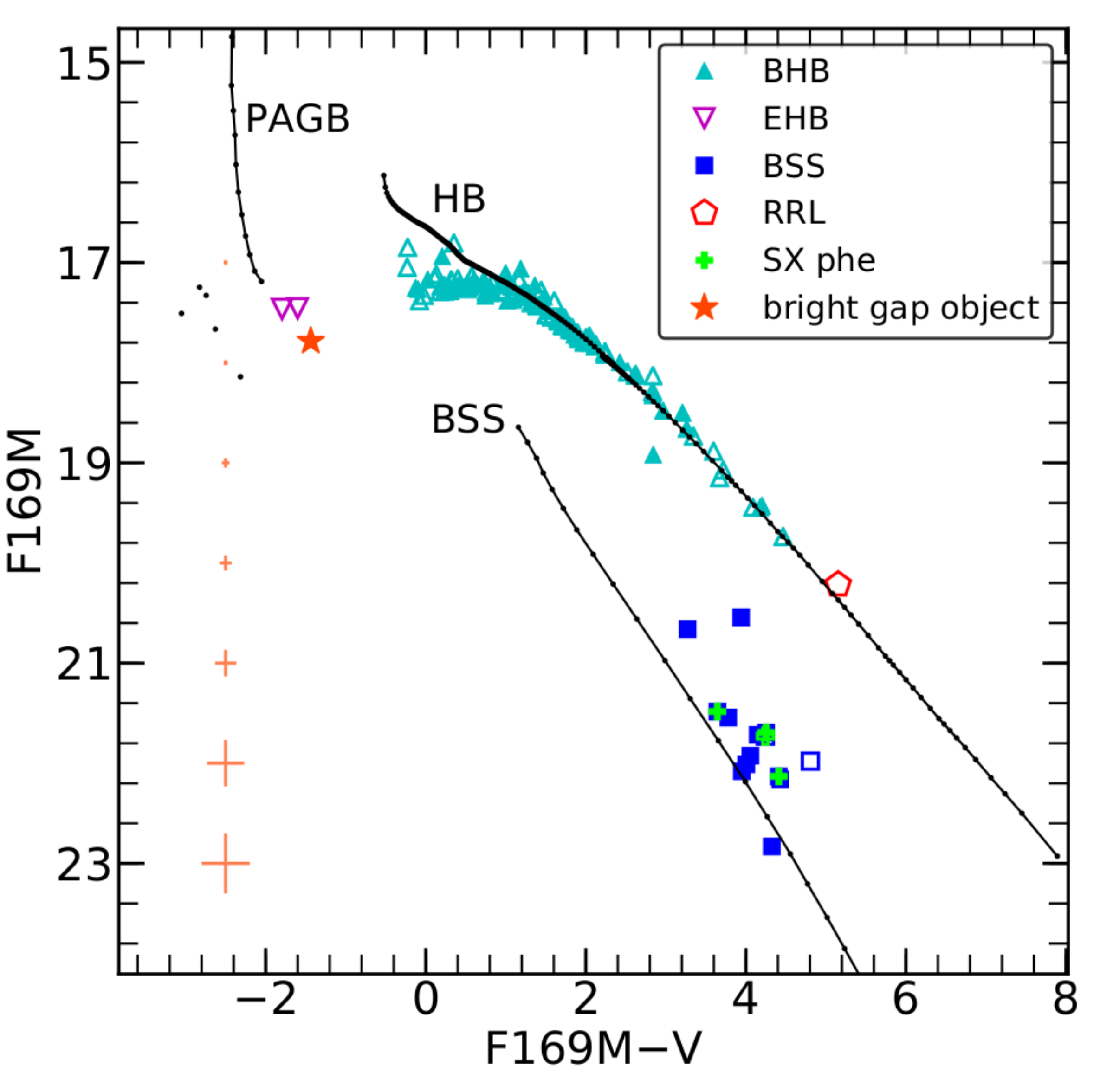}
\caption{F169M$-$V vs F169M CMD. See Figure \ref{caf2_v_cmd} for details.} 
\label{sapph_v_cmd}
\end{center}
\end{figure}

We have cross-matched UVIT data with the HST-ACS and the ground data using TOPCAT \citep{2011topcat}. The CMDs obtained after cross-matching the data in different filters of UVIT are shown in Figures \ref{caf2_v_cmd}-\ref{nuv_cmd}, where the filled symbols are the UVIT cross-matched HST detections and the open symbols are the UVIT cross-matched ground detections. In each figure, we have identified various stellar populations based on their position in the UV CMDs and verified by plotting them in the optical CMDs which are described in the following subsections. The identified populations are shown with same symbols in Figures \ref{caf2_v_cmd}-\ref{nuv_cmd}. We have detected BHB, EHB and BSS population in the FUV CMDs which are marked and shown in Figures \ref{caf2_v_cmd}-\ref{caf2_n2_cmd}. The N279N$-$V versus N279N CMD is shown in Figure \ref{nuv_cmd}, where in addition to the BHB, EHB and BSS population, we have also detected the Main Sequence (MS) and the Red Giant Branch (RGB) stars. These populations are not hot enough to emit in FUV, hence are not detected in the FUV CMDs. 

The CMDs are over-plotted with a BaSTI isochrone \citep{Pietrinferni2004} of 12.6 Gyr \citep{Kaiser2017} and [Fe/H] = $-$1.28 \citep{Carretta2009} adopting a distance modulus of 14.84 \citep{Bella2001}. For generating this isochrone, we have used the Flexible Stellar Population Synthesis (FSPS) model \citep{Conroy2009, Conroy2010} to convolve the BaSTI models with the UVIT filter effective area curves and obtained the model magnitudes in UVIT filters (AB system). The advantage of FSPS Model is that, it also generates the model tracks for HBs, PAGBs \citep{wood1994} and BSSs along with the MS and RGB tracks, which are marked in Figures \ref{caf2_v_cmd}-\ref{nuv_cmd}. The code generates the BSS track that uniformly populates the region from 0.5 magnitude above the main sequence turn-off (MSTO) to 2.5 magnitudes brighter than the turn-off that is primarily based on observations.    

\subsection{Blue Horizontal Branch (BHB)}\label{cmd_bhb}
\begin{figure}
\begin{center}
\includegraphics[scale=0.34]{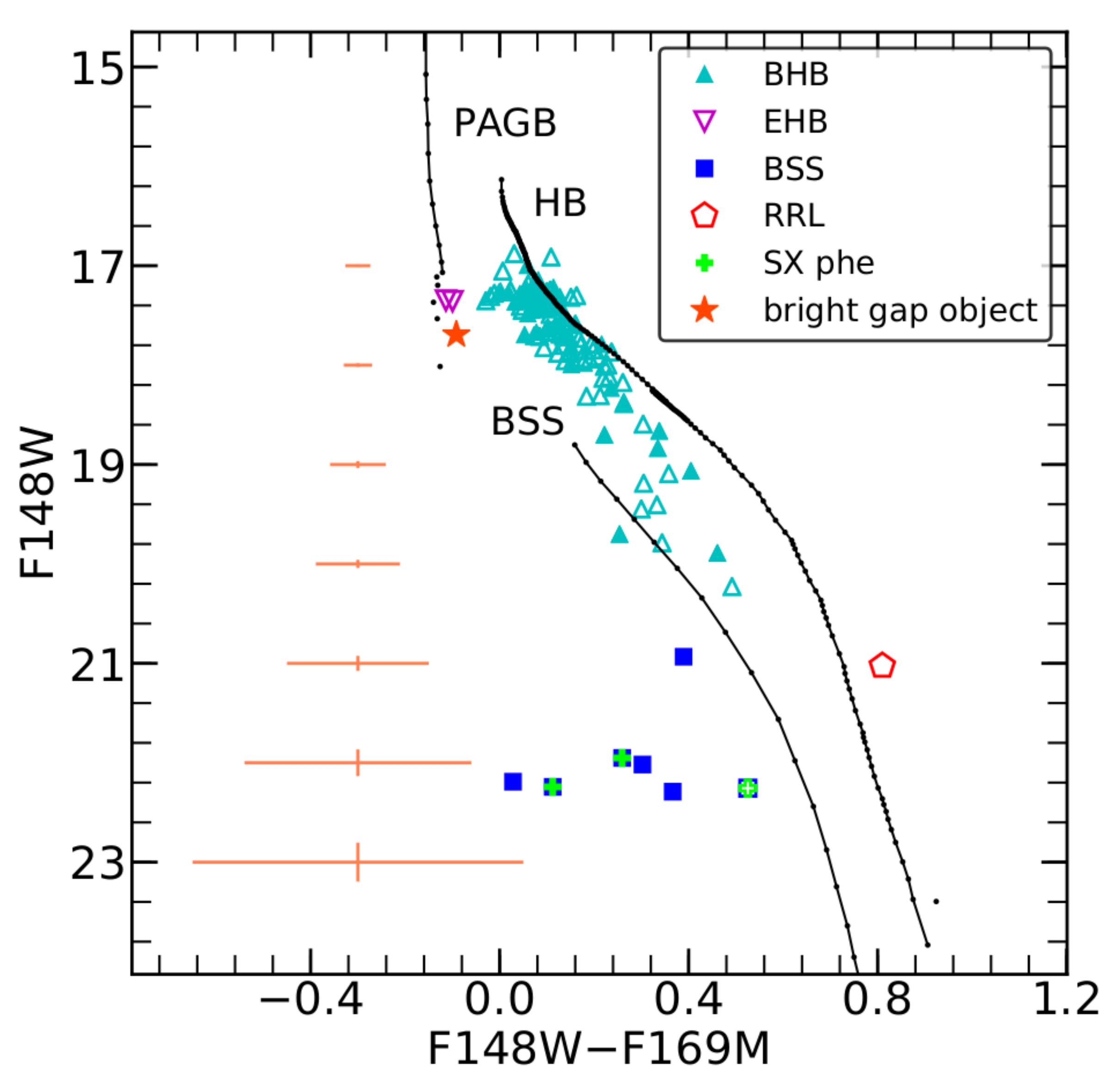}
\caption{F148W$-$F169M vs F148W CMD. See Figure \ref{caf2_v_cmd} for details.}
\label{caf2_sapph_cmd}
\end{center}
\end{figure}

The HB population shown in FUV CMDs in Figures \ref{caf2_v_cmd}-\ref{caf2_n2_cmd} (cyan triangles) mainly consists of BHB stars. In total, we have detected 119 BHB sources in the FUV filters of UVIT and 124 BHB sources in the N279N filter of UVIT with 43 of them falling under the FOV of HST-ACS. The BHBs form a diagonal sequence that spans about 3-5 mag in colour in the UV CMDs (Figure \ref{caf2_v_cmd}, \ref{sapph_v_cmd} and \ref{caf2_n2_cmd}) thus, lifting the degeneracy in the V$-$I colour of the BHB stars in the optical CMD (left panel of Figure \ref{caf2_v_cmd}). The BHB distribution in the FUV vs V CMDs (Figure \ref{caf2_v_cmd} and \ref{sapph_v_cmd}) fits well with the isochrone till F148W $\sim$ 17.4, F148W$-$V $\sim$ 1.2 and F169M $\sim$ 17.3, F169M$-$V $\sim$ 1.2. We note that, stars bluer than FUV$-$V $\sim$ 1.2 form a horizontal sequence to appear like a plateau in the FUV magnitude. This plateau is apparent as the BHB stars are $\sim$ 0.3 mag fainter than the isochrone. This could be attributed to the onset of diffusion which is discussed in Section \ref{discus_hb}. We have detected 4 BHBs in F148W$-$V vs F148W CMD with F148W$-$V $<$ 0.5 which are brighter in F148W magnitude than the rest of the BHB stars. These are classified as AGB-manqu\'e (AGBM) stars by \cite{Schiavon2012}. AGBM stars are those HB stars that after the core-He exhaustion do not have enough envelope masses to evolve to the AGB phases and thus end up with low luminosities as compared to the PAGB stars.

Moving to the FUV CMD which is shown in Figure \ref{caf2_sapph_cmd}, the BHB stars with F148W $>$ 18.5 show a maximum shift of $\sim$ 0.3 mag in F148W$-$F169M colour from the model isochrone which are within the 3$\sigma$ photometric errors. We find a gap in the distribution of BHB stars at F148W $\sim$ 17.5 and F148W$-$F169M $\sim$ 0.08 where, the stars brighter than this magnitude are all bunched together into a group.

The FUV$-$NUV vs FUV CMDs are shown in Figure \ref{caf2_n2_cmd}. Here, we detect a plateau among the BHB stars at F148W $>$ 17.5, F148W$-$N279N $\sim$ 0.3 (upper panel) and F169M $>$ 17.4, F169M$-$N279N $\sim$ 0.3 (lower panel), where the stars bluer than this location deviate from the normal BHB diagonal sequence similar to the FUV vs V CMDs (Figure \ref{caf2_v_cmd} and \ref{sapph_v_cmd}). In addition to this, we also notice that the BHBs fainter than the above mentioned FUV magnitude and colour, deviate from the model isochrone as compared to the FUV vs V CMDs, where they were in good agreement with the isochrone. As N279N is a narrow band filter centred around Mg II (2808 $\mathrm{\AA}$), it captures the Mg abundance variations within the BHB stars. \cite{Moehler2014} derived the Mg abundances of 51 BHB stars in NGC 288 based on the observations obtained from the medium resolution FLAMES-GIRAFFE spectrograph. A closer look and comparison of the UVIT observations with their spectroscopic observations reveal that the redder BHB stars have a higher Mg abundance as compared to the bluer BHB stars. The N279N mag of the BHB stars are fainter than the isochrone which we notice in the NUV vs V CMD (right panel of Figure \ref{nuv_cmd}). Thus, the mismatch between the observed BHB distribution and the isochrone could be due to the difference between the assumed Mg abundance of the model with the observed abundance. The temperature distribution and identification of the gaps in the BHB distribution are described in Section \ref{temp_bhb}. 

\subsubsection{RR Lyrae Variables}
\begin{figure}
\begin{center}
\includegraphics[scale=0.34]{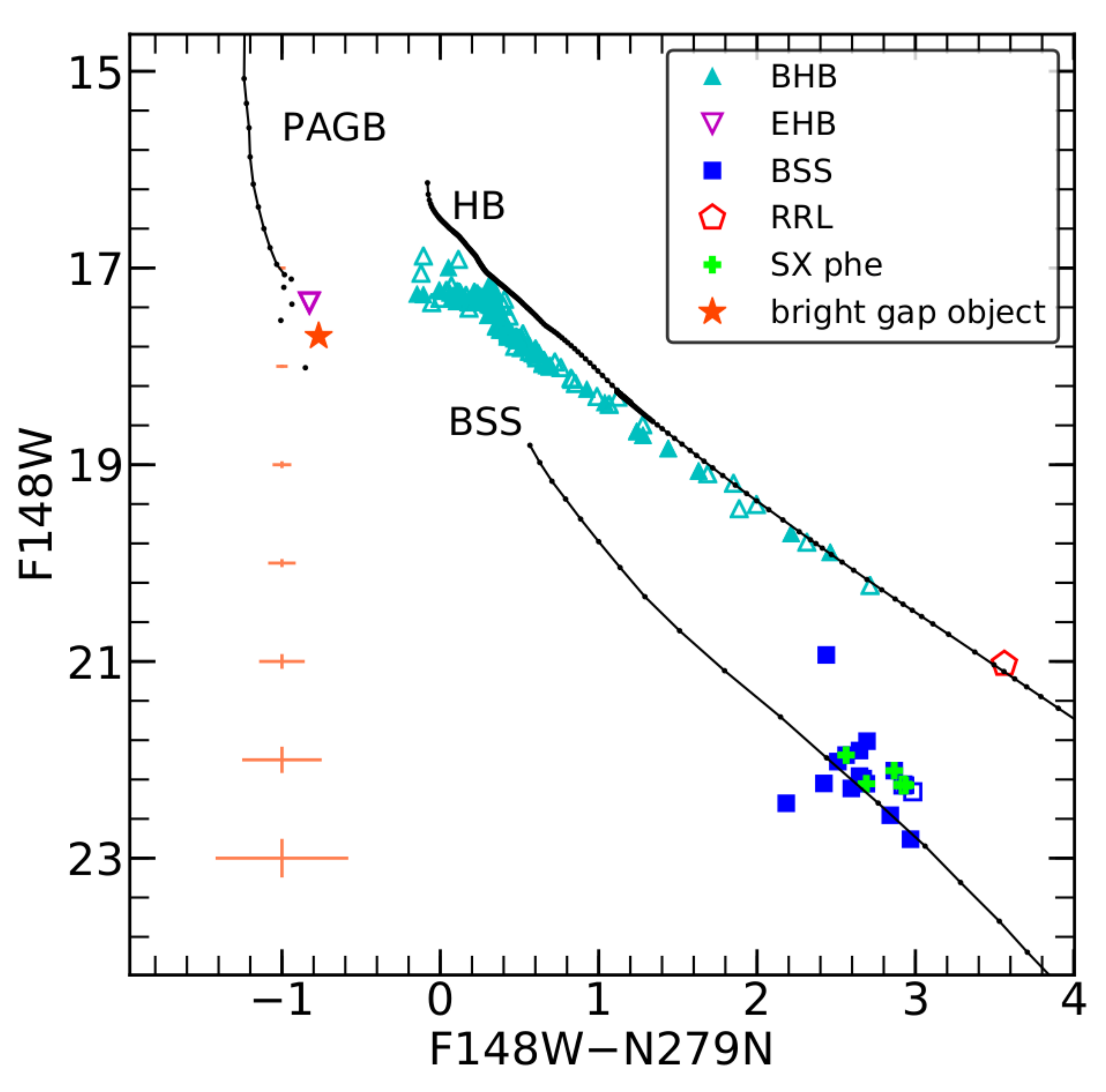}\\
\includegraphics[scale=0.34]{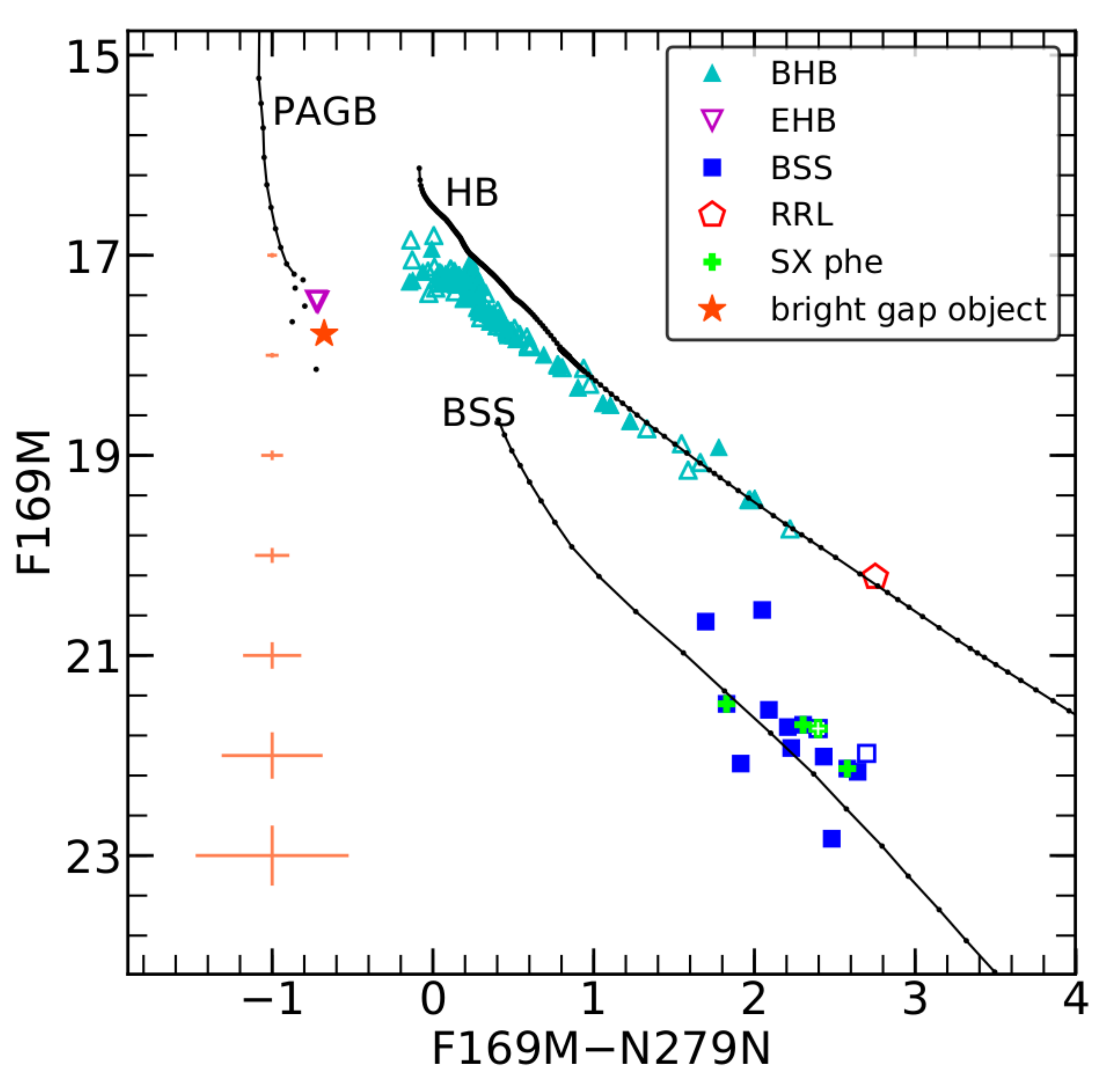}
\caption{F148W$-$N279N vs F148W CMD (upper panel) and F169M$-$N279N CMD vs F169M CMD (lower panel). See Figure \ref{caf2_v_cmd} for details.}
\label{caf2_n2_cmd}
\end{center}
\end{figure}

\cite{Kaluzny1996} and  \cite{Ferro2013} found 2 RR Lyrae variables based on the V band light curves obtained from the ground-based observations and derived their properties. We have detected these two RR Lyrae variables in the N279N filter of UVIT and only one RR Lyrae in the FUV filters of UVIT, whose locations in the optical and UV CMDs are shown in Figures \ref{caf2_v_cmd}-\ref{nuv_cmd} as red pentagons. As expected, the RR Lyrae variables are located at the red end of the BHB distribution in all the UV CMDs. The FUV bright RR Lyrae is of RRc type with a shorter period as compared to the NUV bright RR Lyrae of RRab type \citep{Ferro2013}.

\subsection{Extreme Horizontal Branch (EHB)}
EHB stars are defined as the hottest BHB stars with effective temperatures greater than 20,000 K that undergo severe mass loss during the RGB phase \citep{Heber1986}. \cite{Kaluzny1996} reported the presence of three likely hot sub-dwarfs located in the extension of BHB in the optical CMD. In the FUV CMD of GC NGC 2808, \cite{Dalessandro2011} have shown the location of different HB stars, where we can clearly notice a large population of EHBs bluer than the BHB stars. Considering the definition adopted by \cite{Dalessandro2011}, and keeping the HB and PAGB model isochrone as reference for the selection of EHB stars in the UV CMD, we have found 2 potential EHB candidates which are shown in Figures \ref{caf2_v_cmd}-\ref{nuv_cmd} (open magenta triangles). These stars are located in the expected EHB region in both the optical and UV CMDs having similar FUV magnitude as the BHB stars. GALEX has also detected these stars which are located in the blue extension of BHB in the FUV$-$NUV vs FUV CMD of NGC 288 \citep{Schiavon2012}. These EHB candidates are two of the three sub-dwarfs as reported by \cite{Kaluzny1996}. 
The SEDs of the EHB candidates are discussed in Section \ref{temp_ehb}. 

\subsection{Gap Objects}
Gap objects are the stars which are located between the MS and white dwarf (WD) sequence in the optical and UV CMDs. The CVs are expected to lie in this region as pointed out by \cite{Haurberg2010}, \cite{Dieball2010} and \cite{Dieball2017} in their study of M15, M80 and NGC 6397 respectively. We have found one bright gap object which is marked as orange star in Figures \ref{caf2_v_cmd}-\ref{nuv_cmd}. This object which is located in the gap region in optical CMD becomes very bright in UV CMDs with its location being close to that of EHB stars. In the UV CMDs, it is fainter by $\sim$ 0.35 mag with a similar colour as compared to the EHB stars. We checked for its variability with the known catalogue of X-ray sources and CVs in this cluster available from the previous study by \cite{Kong2006}, but we did not found any match for this object. The SED of this object is described in Section \ref{temp_ehb}.

\subsection{Main Sequence and Red Giant Branch}
\begin{figure*}
\begin{center}
\includegraphics[scale=0.35]{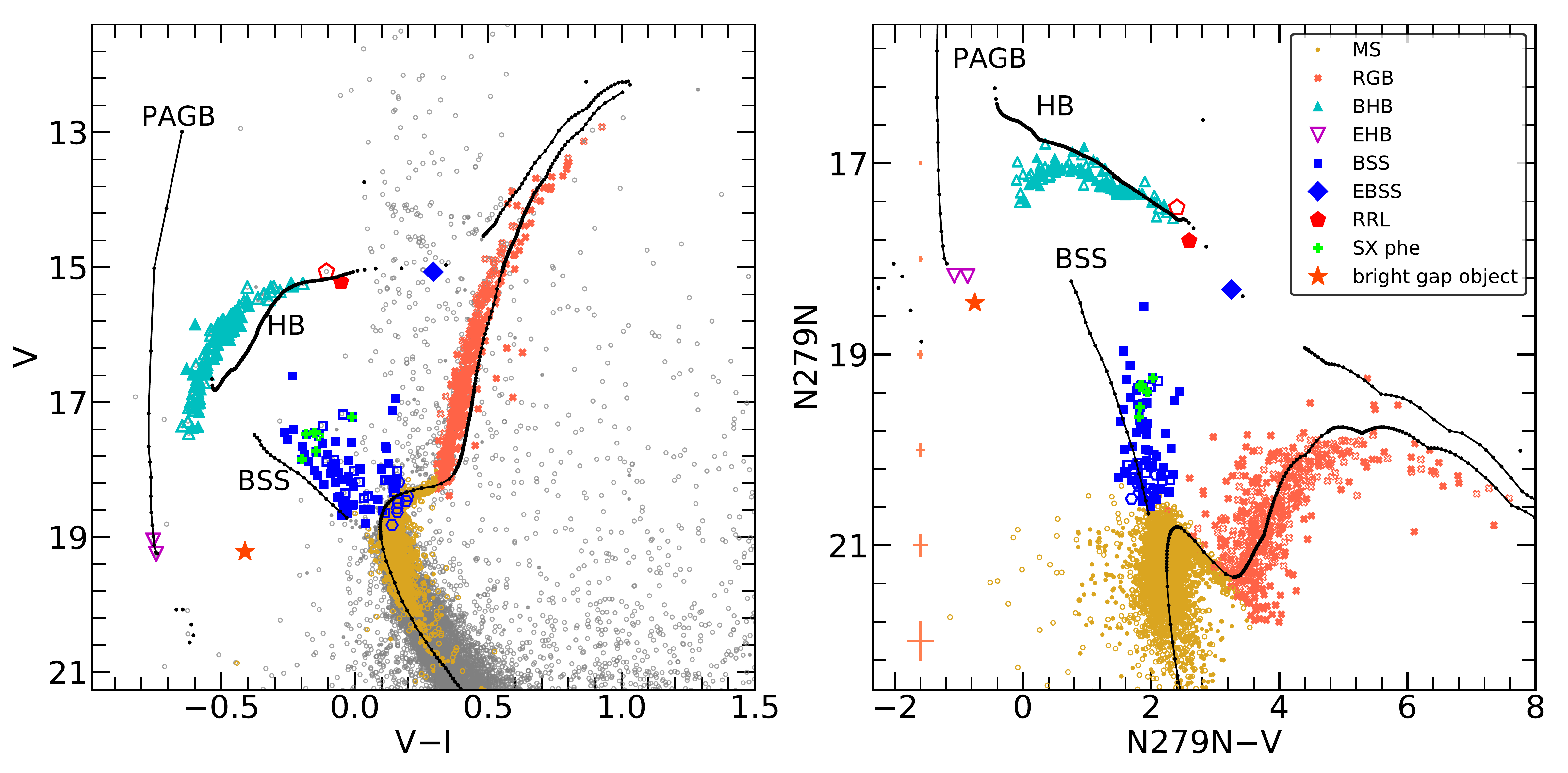}
\caption{N279N$-$V vs N279N CMD (right panel) and corresponding optical CMD (left panel). In addition to the detection of other stellar populations, we also detect main sequence marked as MS (yellow dots) and red giant branch marked as RGB (red cross) in the N279N filter of UVIT. See Figure \ref{caf2_v_cmd} for more details.}
\label{nuv_cmd}
\end{center}
\end{figure*}

The MS extends $\sim$ 2.0 mag below the turn-off till $\sim$ 22 mag in N279N (yellow dots in Figure \ref{nuv_cmd}) with a width of $\sim$ 1 mag. We see a large scatter in the MS in NUV CMD for stars with N279N $>$ 21, partially due to photometric errors. The Sub-giant Branch (SGB) is narrow as compared to the RGB and MS for the same photometric errors in N279N mag. The BaSTI isochrone shown in Figure \ref{nuv_cmd} visually fits well with the observed MS, SGB and RGB distribution. A number of faint sources are detected bluer than the MS, which could be gap objects or MS stars with NUV excess.  

\subsection{Blue Straggler Stars (BSSs)}\label{bss_phot}
We present the UV CMDs of the BSSs covering the entire cluster region in this study. We have detected 17 BSS candidates in F148W filter and 14 BSS candidates in F169M filter of UVIT whose location in the FUV CMDs are shown in Figures \ref{caf2_v_cmd}-\ref{caf2_n2_cmd}. The BSSs stretch $\sim$ 2 mag in the FUV CMDs. In Figure \ref{caf2_sapph_cmd}, we see a spread of $\sim$ 0.5 mag in F148W$-$F169M colour with F148W magnitude being similar for all 6 BSSs. This spread is within the photometric errors.

We have detected 78 BSS candidates in N279N$-$V vs N279N CMD marked as blue squares in Figure \ref{nuv_cmd} (right panel) among which 7 of them are new candidates (marked as blue hexagons). These 7 new BSS candidates are located in the BSS region in the NUV CMD (marked as blue stars) whereas, they are near the MSTO in the optical CMD. Based on the HST-UV observations, \cite{Raso2017} showed that the NUV CMDs are more suitable for the identification of BSS candidates as compared to the optical CMDs where the faint BSSs remain hidden near the MSTO. 

We notice that BSSs have similar magnitude and colour range in optical and NUV CMDs. We have also detected a star located in the region of red HB stars which is shown as blue diamond in Figure \ref{nuv_cmd}. \cite{Bellazzini2001} also found this object to be located in the red HB region and classified it as an Evolved BSS (EBSS). The EBSS is not detected in the FUV filters of UVIT. The parameters of the BSS derived from the photometry and SEDs are described in the Section \ref{param_bss}.

\subsubsection{SX Phoenicis Variables (SX Phes)}
SX Phe variables are short period pulsating variables found in the location of BSSs in the optical CMDs of GCs. In total, there are 8 known SX Phe variables in this cluster from the previous studies \citep{Kaluzny1996, Kaluzny1997, Ferro2013, Martinazzi2015}. We have cross-matched UVIT data with the coordinates of the known SX Phe variables to obtain their UV magnitudes in three different filters of UVIT. We have detected 5 of them in F148W, 4 in F169M and 6 in N279N filters of UVIT which are shown as green plus symbols in Figures \ref{caf2_v_cmd}-\ref{nuv_cmd}. The three SX Phe variables, V8, V11, and V12 also show a spread in F148W$-$F169M colour like BSSs (Figure \ref{caf2_sapph_cmd}) but are within the photometric errors. In N279N$-$V vs N279N CMD (right panel of Figure \ref{nuv_cmd}) we see that the variables stand out clearly from the MSTO stars as compared to the optical CMD. The variables are located in the region of BSSs with N279N $<$ 20 that spans $\sim$ 0.4 mag in N279N and $\sim$ 0.2 mag in N279N$-$V colour. 

\begin{table*}
\centering
\caption{A sample catalogue (3 BHBs) of the UV detected possible member stars \citep{Gaia2018b} is presented here. The full catalogue is available in electronic format. Column 1 corresponds to our Star ID, columns 2 \& 3 list the RA and Dec of the stars, columns 4 to 9 give the UVIT magnitudes and errors in F148W, F169M and N279N filters respectively, columns 10 and 11 give the optical magnitude and colour corresponding to V and (V$-$I) from the HST-ACS \citep{Sarajedini2007} and ground data (Peter Stetson, private comm.) as mentioned in column 12, column 13 gives the radial distance from the cluster centre (r) and columns 14 to 17 give the proper motions in RA ($\mu_{\alpha}cos\delta$) and Dec ($\mu_{\delta}$) with corresponding errors available from GAIA DR2 \citep{Gaia2018b}. Note that the magnitudes and colours (AB system) are not corrected for extinction and reddening respectively.}
\resizebox{170mm}{!}{
\begin{tabular}{ccccccccccccccccc}
\hline
\hline
Star ID & RA (J2000) & Dec (J2000) & F148W & err1 & F169M & err2 & N279N & err3 & V &\\ 
& [h m s] & [$\degr$ $\arcmin$ $\arcsec$] & [mag] & [mag] & [mag] & [mag] & [mag] & [mag] & [mag] &  \\\hline
V$-$I & Optical data & r & pmra & pmra$\_$err & pmdec & pmdec$\_$err\\
$[$mag$]$ & & [$'$] & [mas$/$yr] & [mas$/$yr] & [mas$/$yr] & [mas$/$yr] \\\hline
BHB1	&	00	52	48.31	&	-26	32	58.21	&	17.880	&	0.019	&	17.768	&	0.026	&	17.414	&	0.028	&	15.935	&\\	-0.552	&	HST	&	2.43	&	4.391	&	0.106	&	-5.440	&	0.078	\\
BHB2	&	00	52	48.59	&	-26	33	17.74	&	17.532	&	0.017	&	17.472	&	0.025	&	17.380	&	0.038	&	17.149	&\\	-0.552	&	HST	&	2.11	&	4.366	&	0.272	&	-5.572	&	0.149	\\
BHB3	&	00	52	45.47	&	-26	33	22.72	&	17.558	&	0.015	&	17.418	&	0.023	&	17.285	&	0.028	&	16.677	&\\	-0.561	&	HST	&	2.06	&	4.850	&	0.184	&	-6.328	&	0.156	\\\hline
\label{uvit_tab}
\end{tabular}
}
\end{table*}

\subsection{Cluster membership from GAIA DR2}
In order to check the cluster membership of BHBs, EHBs and BSSs detected by UVIT, we have used the GAIA DR2 catalogue of NGC 288 \citep{Gaia2018b}, where they have provided the list of possible members in this cluster based on the proper motions. The UVIT data was cross-matched with GAIA DR2 to select the UV bright cluster members. The final sample of members comprises of 110 BHB stars detected in FUV filters and 115 in NUV filter. We have excluded 9 BHB stars that are non-members that are located within the radius $r \sim 2.5'$ from the cluster centre. Similarly, we are left with 15 FUV detected BSSs and 68 NUV detected BSSs which are cluster members. We have excluded 10 BSSs that are located at a radius beyond 2$'$ from the cluster centre as they were found to be non-members. Out of 68 BSSs, there are 12 BSSs which do not have proper motion data available in the catalogue though these are observed by GAIA. We assumed them to be cluster members as most of them ($\sim 75\%$) lie inside the half-light radius of the cluster i.e. in the crowded regions. The typical uncertainties in the proper motions for BHBs and BSSs are $\sim$ 0.16 and 0.21 mas/yr respectively. The 2 EHBs and the bright gap object are also possible cluster members but have larger proper motion errors ($\sim$ 0.7-1 mas/yr) as compared to the BHBs and BSSs. A sample catalogue of UVIT detected members along with the proper motions from GAIA DR2 is given in Table \ref{uvit_tab}. The rest of the analysis is based on this sample.

\section{Temperature Distribution of BHB stars}\label{temp_bhb}
We have determined the temperature of the detected BHB stars using two different methods and compared them with the spectroscopic estimates of \cite{Moehler2014} which are described below:  

\begin{figure}
\begin{center}
\includegraphics[scale=0.34]{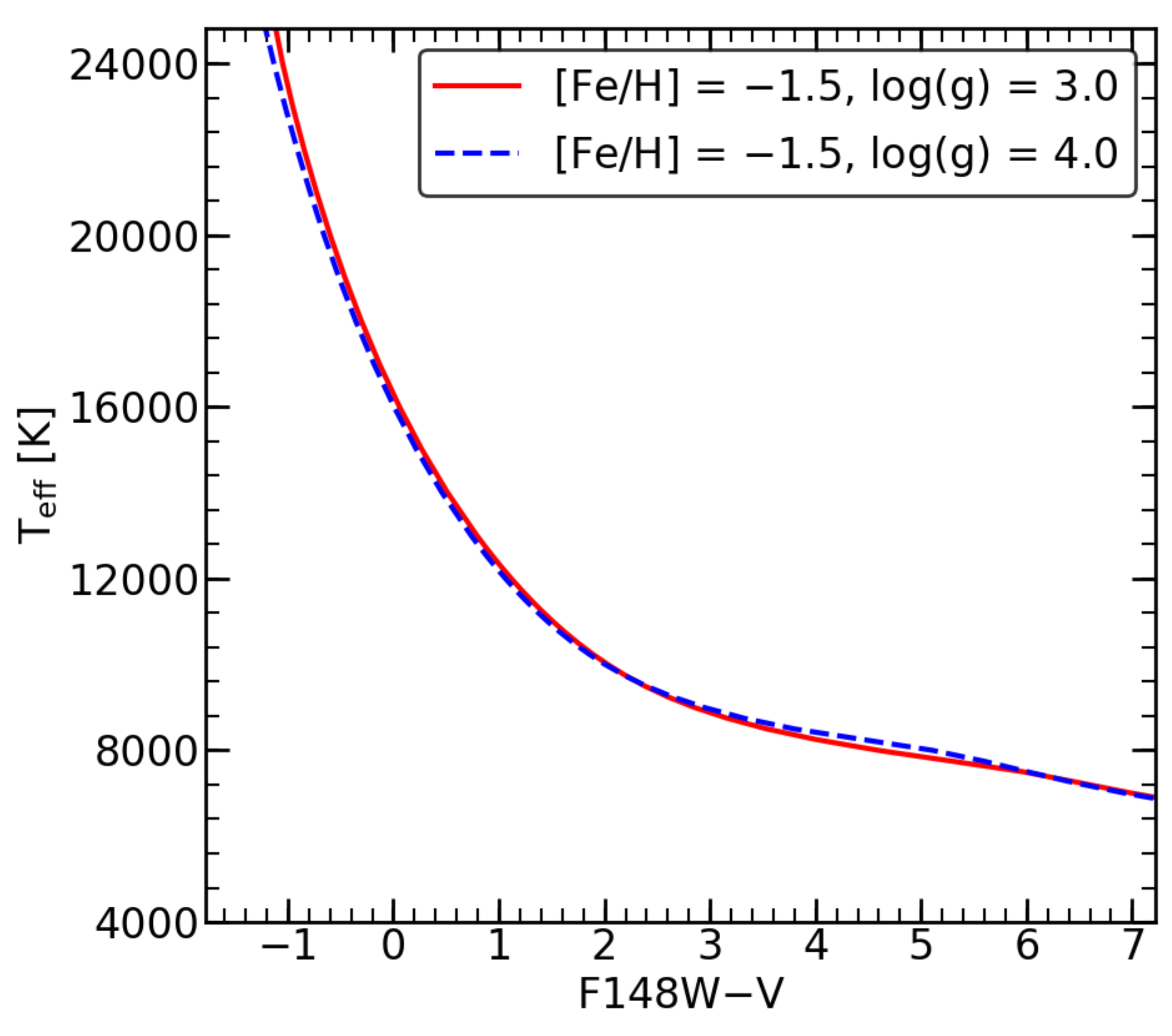}
\includegraphics[scale=0.34]{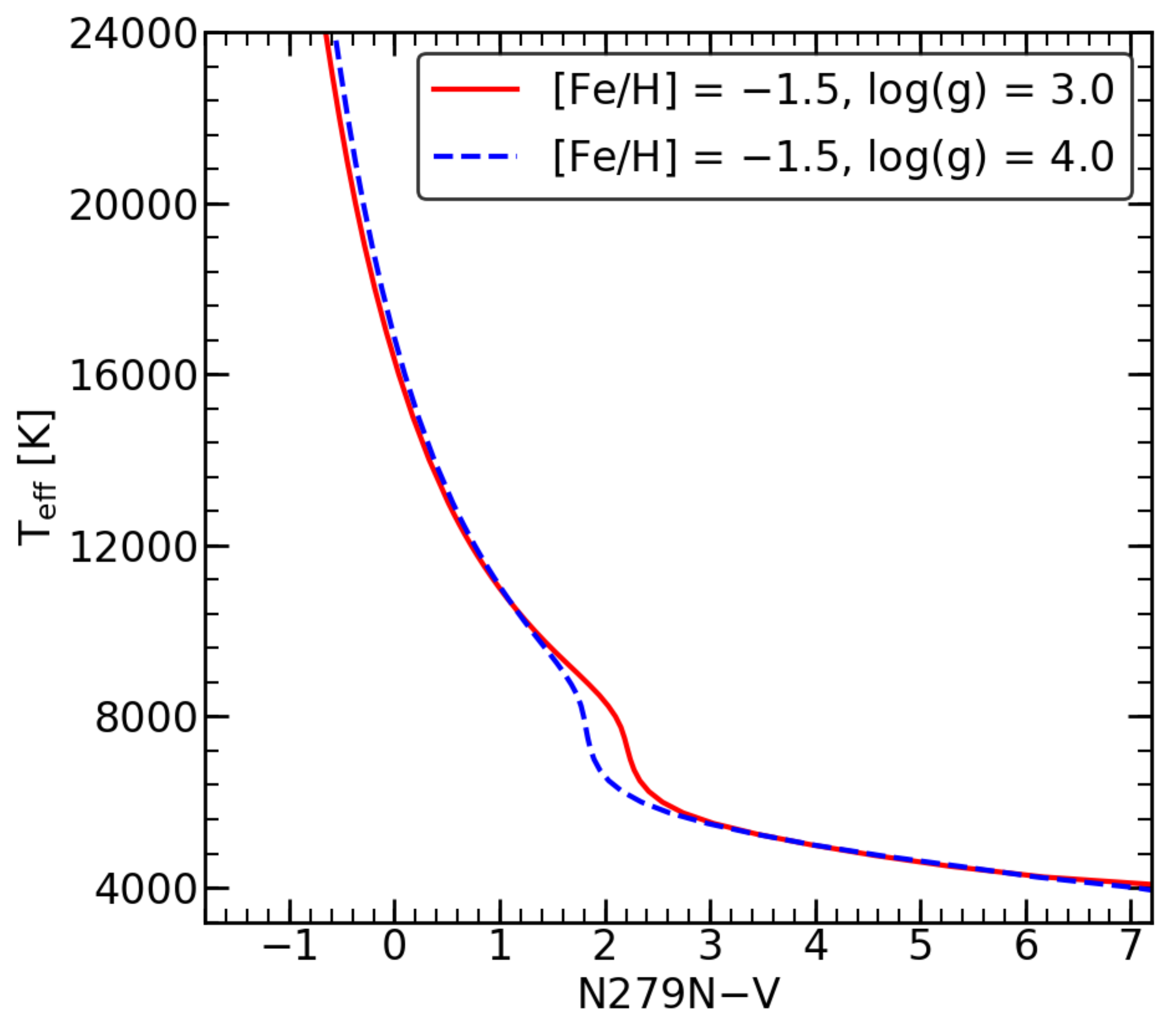}
\caption{Variation of temperature ($T_{eff}$) with theoretical colours F148W$-$V (upper) and N279N$-$V (lower) generated by convolving UVIT filter effective area curves with Kurucz stellar atmospheric models \citep{Castelli1997} for an assumed metallicity of [Fe/H] = $-$1.5 and log(g) values of 3.0 and 4.0.}
\label{model}
\end{center}
\end{figure}

\subsection{Temperature from colour - $T_{eff}$ relation}\label{temp_color}
We have estimated the temperature of the BHB stars by comparing the dereddened UV colours with the theoretical colours of various temperatures. In order to do so, we have convolved Kurucz stellar atmospheric models \citep{Castelli1997} with the UVIT filter effective area curves, and estimated the theoretical colours for different temperatures for a metallicity [Fe/H]= $-$1.5 closest to the cluster metallicity \citep{Carretta2009} and for a primordial helium abundance of Y = 0.248. We have selected model spectra within 3.0 $\leq$ log(g) $\leq$ 4.0 and 4,000 K $\leq T_{eff} \leq$ 24,000 K to derive the theoretical UV colours as the temperature of the BHB stars are expected to be within this range. The colour - $T_{eff}$ relation for a log(g) of 3.0 and 4.0 and for two UVIT filter combinations (F148W$-$V and N279N$-$V) are shown in Figure \ref{model} in upper and lower panels respectively. It is clear from this figure that the F148W$-$V colour is more sensitive to $T_{eff}$ variations as compared to N279N$-$V. In both the colours, the effect of surface gravity is visible below $\sim$ 8000 K. This is in accord with the study of the HB temperature distribution of M15 by \cite{Lagioia2015}, where they found that the HB tracks of different masses show dependency of surface gravity below $\sim$ 8000 K for the NUV$-$V colour. We have also found that the metallicity variations do not significantly affect the colour - $T_{eff}$ relation which is in agreement with \cite{Lagioia2015}. Typical photometric errors of the BHB stars in the colour F148W$-$V and N279N$-$V are $\sigma \sim$ 0.02 and 0.04 which lead to a small error of $\Delta T_{eff} \sim$ 100 K and 170 K respectively.

\begin{figure}
\begin{center}
\includegraphics[scale=0.335]{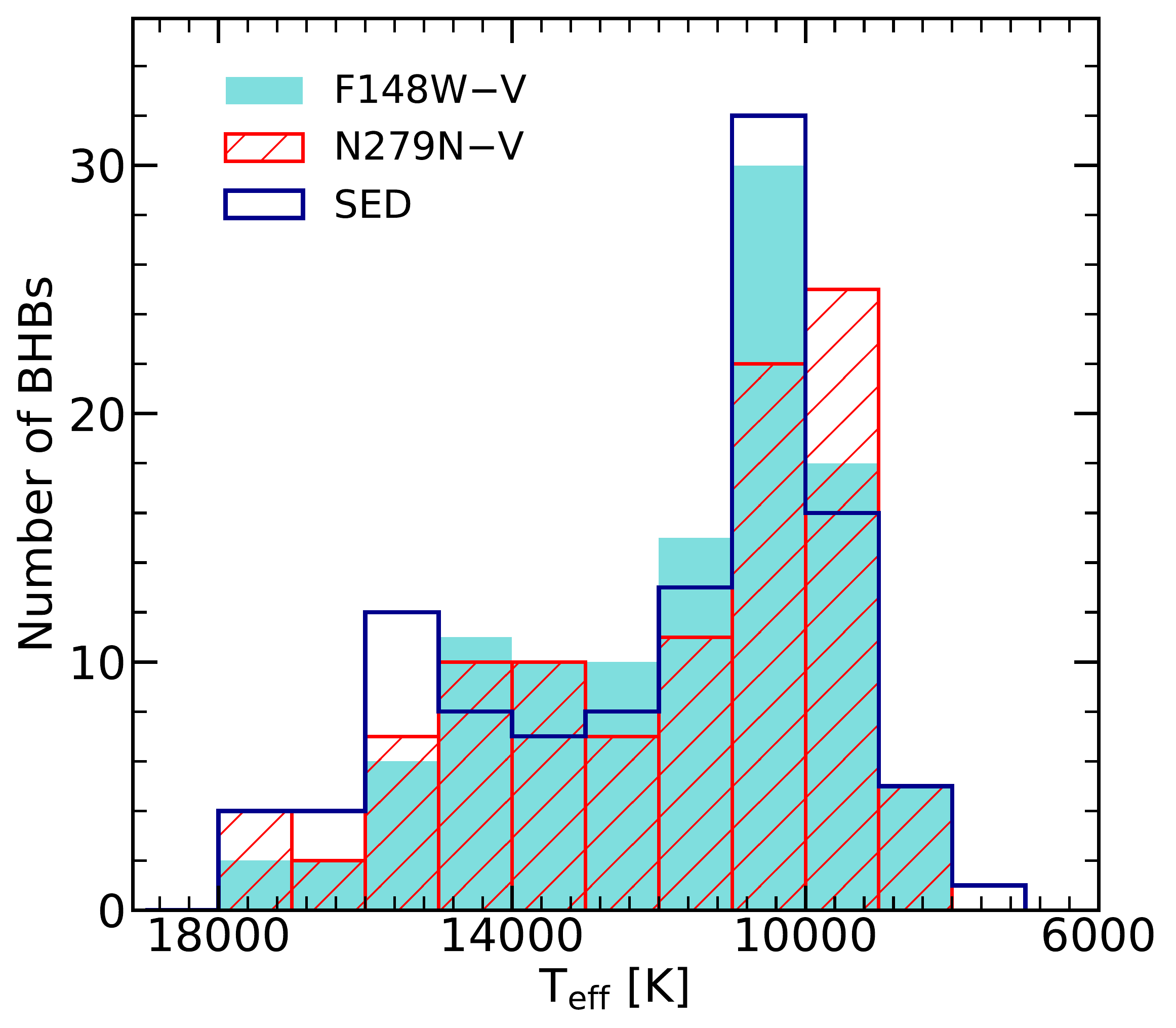}
\caption{Temperature distribution of 109, 103 and 110 BHB stars estimated from F148W$-$V - $T_{eff}$ relation (cyan filled), N279N$-$V - $T_{eff}$ relation (red hatched) and SED fitting (dark blue step) respectively.}
\label{hist_colour}
\end{center}
\end{figure}

\begin{figure}
\begin{center}
\includegraphics[scale=0.34]{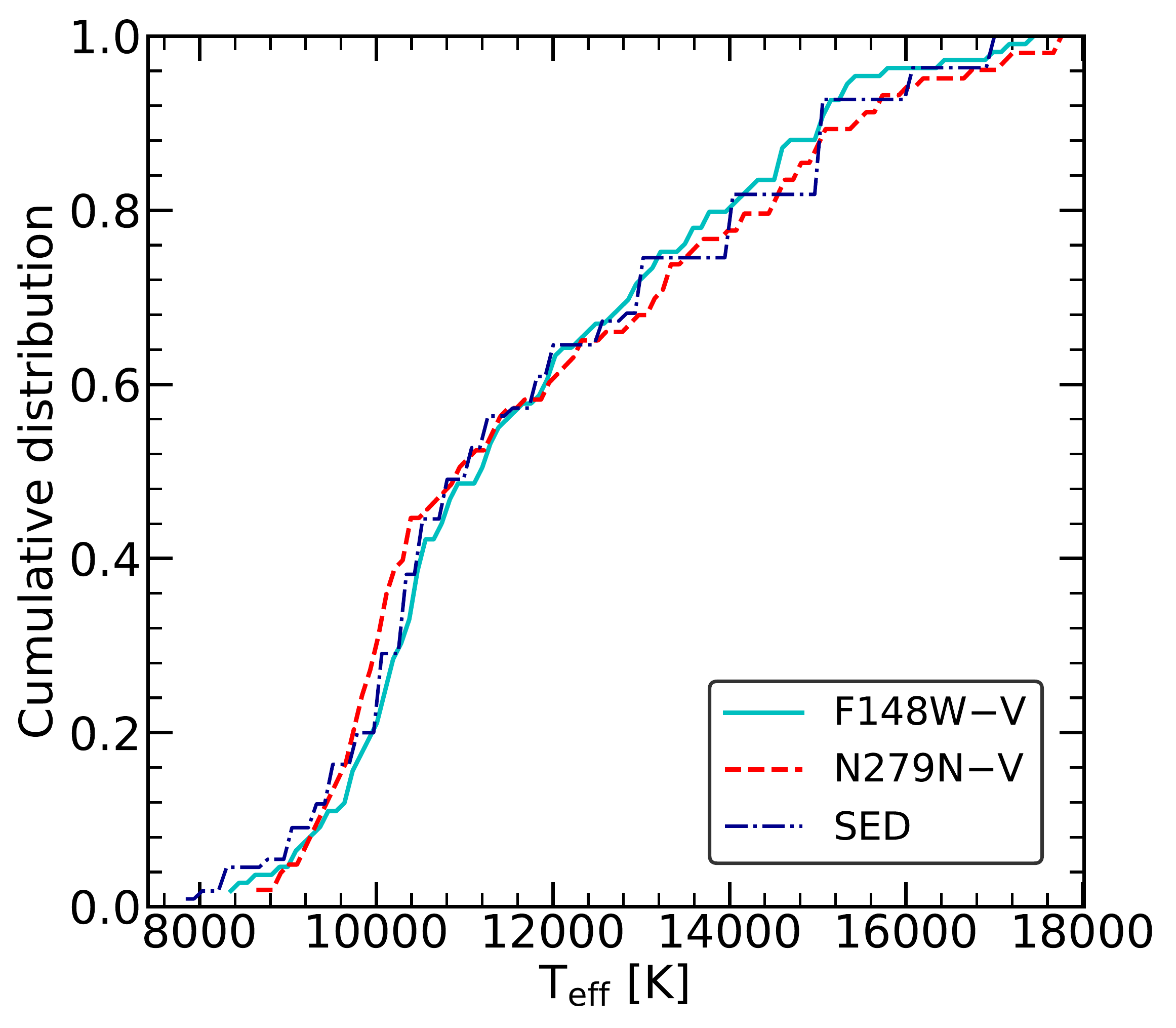}
\caption{Cumulative temperature distribution of the BHB stars obtained from F148W$-$V - $T_{eff}$ (cyan line), N279N$-$V - $T_{eff}$ (red dashed line) and SED fitting (blue dash dotted line).}
\label{cum_hb_tem}
\end{center}
\end{figure}

We have used the dereddened colours for deriving the $T_{eff}$ of the BHBs by doing a cubic interpolation along the theoretical colour - $T_{eff}$ relation.  The BHBs with $T_{eff} < $ 8000 K obtained using the colour - $T_{eff}$ relation were not considered due to the effect of surface gravity as mentioned earlier. The temperature distribution of 109 BHBs obtained from F148W$-$V - $T_{eff}$ relation (cyan filled histogram) and 103 BHBs obtained from N279N$-$V - $T_{eff}$ relation (red hatched histogram) are shown in Figure \ref{hist_colour}. We have identified peaks in the BHB distribution using dual Gaussian fits. The temperature distribution of the BHBs show a main peak located at $T_{eff} \sim$ 10,300 K and another peak at $\sim$ 14,000 K which extends up to $\sim$ 18,000 K. The cumulative temperature distributions of the BHBs obtained from the two different UVIT colours are shown in Figure \ref{cum_hb_tem}. According to the Kolmogorov-Smirnov (K-S) test, the difference between the distributions of effective temperatures obtained from both the UVIT colours is not significant with a p-value of $\sim$ 0.45.

\begin{figure}
\begin{center}
\includegraphics[scale=0.305]{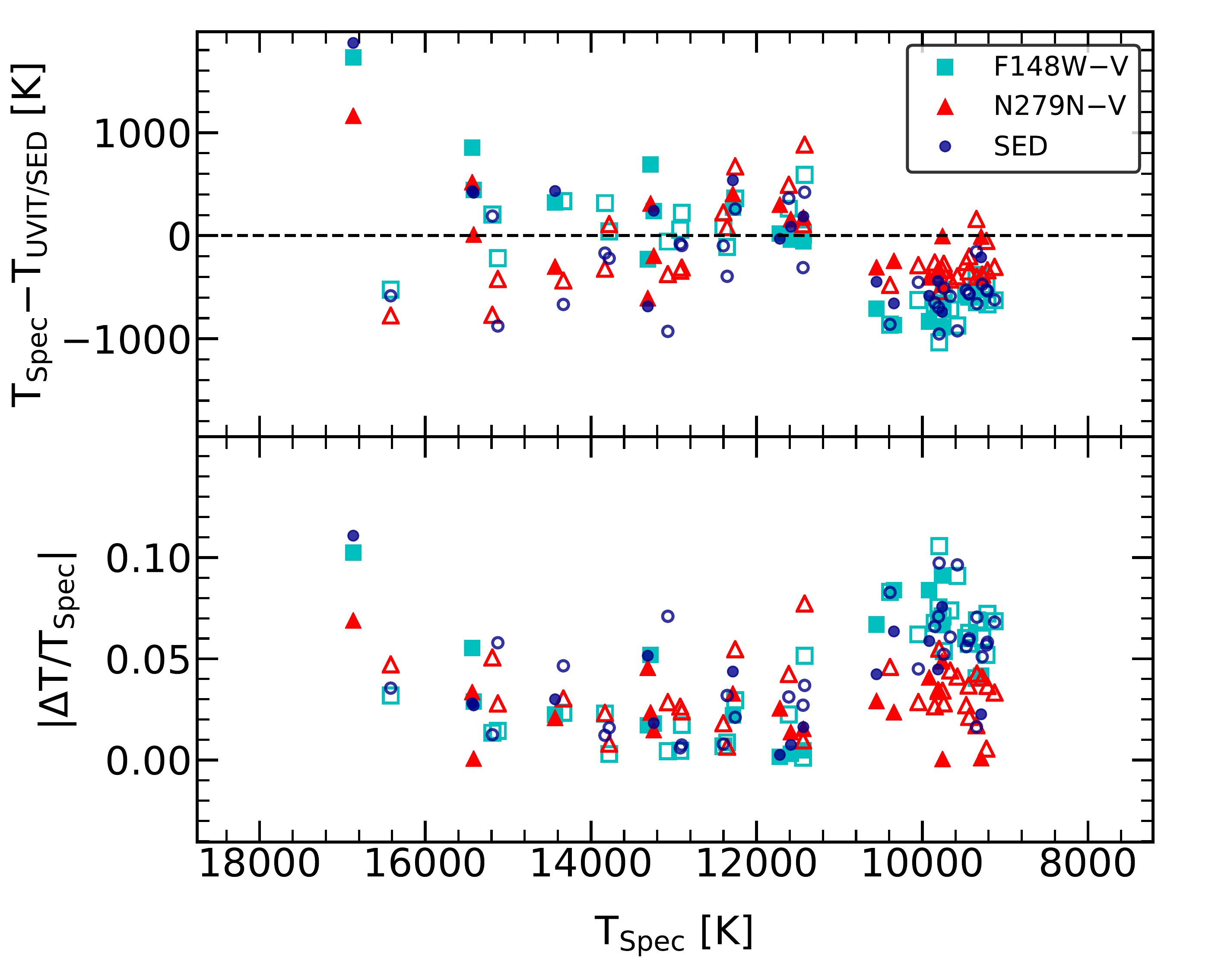}
\caption{Top panel: The difference between the temperatures estimated from spectroscopy (T$_{Spectro}$) (\citep{Moehler2014} and photometry/SED (T$_{UVIT}$) obtained from F148W$-$V - $T_{eff}$ (cyan squares), N279N$-$V - $T_{eff}$ (red triangles) relation and SED fitting (blue circles) respectively.\\
Bottom panel: The absolute value of the fractional differences between the temperatures estimated from spectroscopy and colour - $T_{eff}$/SED fitting as mentioned earlier.
The filled circles are the UVIT cross-matched HST detections and the open circles are UVIT cross-matched ground detections in both the panels.}
\label{spec_phot}
\end{center}
\end{figure}

\subsection{Temperature from SED fitting}\label{temp_sed}
We have obtained the temperature of 110 BHB stars using the python SED fitter tool \citep{Robitaille2007}. For this, we have used Kurucz stellar atmospheric models \cite{Castelli1997} to fit the SEDs of the BHBs stars. The model fluxes are scaled to the observed fluxes by considering a cluster distance of 8.8 Kpc \citep{Bella2001}. We have used the dereddened fluxes obtained from three filters of UVIT (F148W, F169M, N279N), three filters from Ground (B, V, I) to generate the SEDs of 74 BHB stars and three filters of UVIT (F148W, F169M, N279N), two filters of HST (V, I) for the SEDs of 36 BHB stars. The temperature distribution of 110 BHB stars is shown in Figure \ref{hist_colour} (blue histogram). We have found a main peak located at $T_{eff} \sim$ 10,380 K and another peak at $T_{eff} \sim$ 14,600 K using dual Gaussian fits which are similar to those estimated using colour - $T_{eff}$ relation. The temperature range of the BHBs are also similar in both the cases. According to the Kolmogorov-Smirnov (K-S) test as shown in Figure \ref{cum_hb_tem}, the difference between the distributions of effective temperatures obtained from SEDs and F148W$-$V and N279N$-$V colour - T$_{eff}$ relation is not significant with a p-value of $\sim$ 0.65 and 0.55 respectively. This test also suggests that the temperatures of the BHBs estimated using SED fitting are in better agreement with those estimated using F148W$-$V as compared to N279N$-$V colour - T$_{eff}$ relation.

\subsection{Comparison with Spectroscopic data}
As described earlier in Section \ref{cmd_bhb}, \cite{Moehler2014} catalogue consists of $T_{eff}$ ($>$ 9000 K) and log(g) for 51 BHB stars based on spectroscopy. We have used the spectroscopic temperature estimates of BHBs to validate with our temperature estimations using the colour - $T_{eff}$ relation and SED fitting. Our comparisons of temperatures of 51 BHB stars obtained using F148W$-$V - $T_{eff}$, N279N$-$V - $T_{eff}$ relation and SED fitting with the spectroscopic data are shown as cyan squares, red triangles and blue circles respectively in Figure \ref{spec_phot}. The top panel of Figure \ref{spec_phot} shows the difference between the temperatures estimated from spectra and colour - $T_{eff}$ relation/SED fitting whereas the bottom panel shows the fractional difference. The median difference of temperatures are $\sim$ 214 K, 95 K and 30 K for $T_{eff} >$ 11,000 K and $\sim$ $-$666 K, $-$334 K and $-$566 K for $T_{eff} <$ 11,000 K for F148W$-$V, N279N$-$V colour - $T_{eff}$ relation and SED fitting respectively. In the bottom panel of the Figure \ref{spec_phot}, all the BHB stars scatter around $|\Delta T/T_{Spec}|$ = 0.018, 0.026 and 0.03 for $T_{eff} >$  11,000 K and around $|\Delta T/T_{Spec}|$ = 0.068, 0.034, 0.06 for $T_{eff} <$  11,000 K for F148W$-$V, N279N$-$V colour - $T_{eff}$ relation and SED fitting respectively. 

Overall, the temperatures of BHB stars with $T_{eff} >$ 11,000 K obtained from both the methods have better consistency with the spectroscopic estimates. Further, the comparisons of the two methods with \cite{Moehler2014} suggest that SED fitting or F148W$-$V - $T_{eff}$ should be preferred for estimating the temperatures of BHBs with $T_{eff} >$ 11,000 K whereas N279N$-$V - $T_{eff}$ relation for BHBs with temperatures between 8000 K $< T_{eff} <$ 11,000 K.

\subsection{Identification of Gaps}

\begin{figure}
\begin{center}
\includegraphics[scale=0.32]{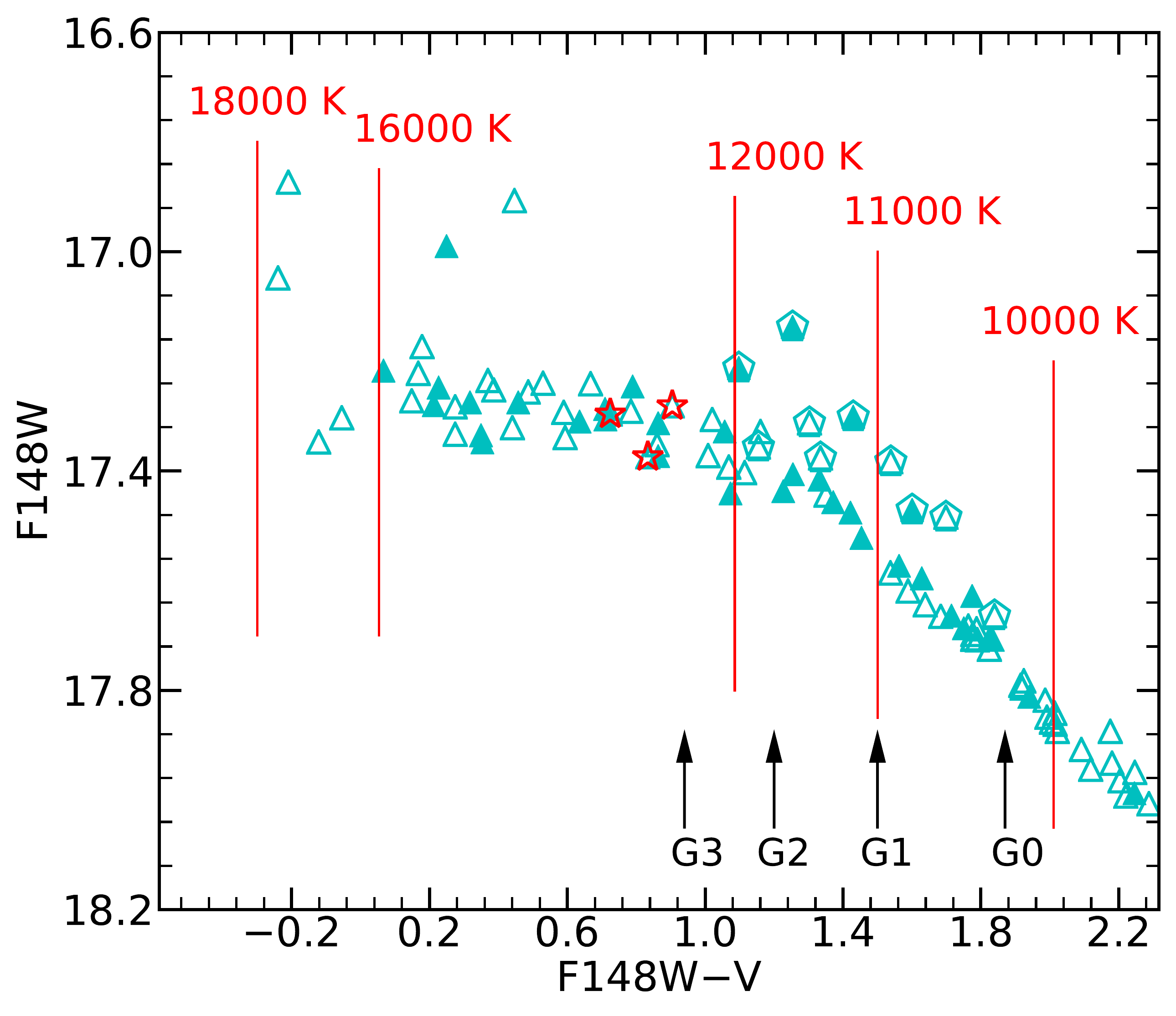}\\ 
\includegraphics[scale=0.33]{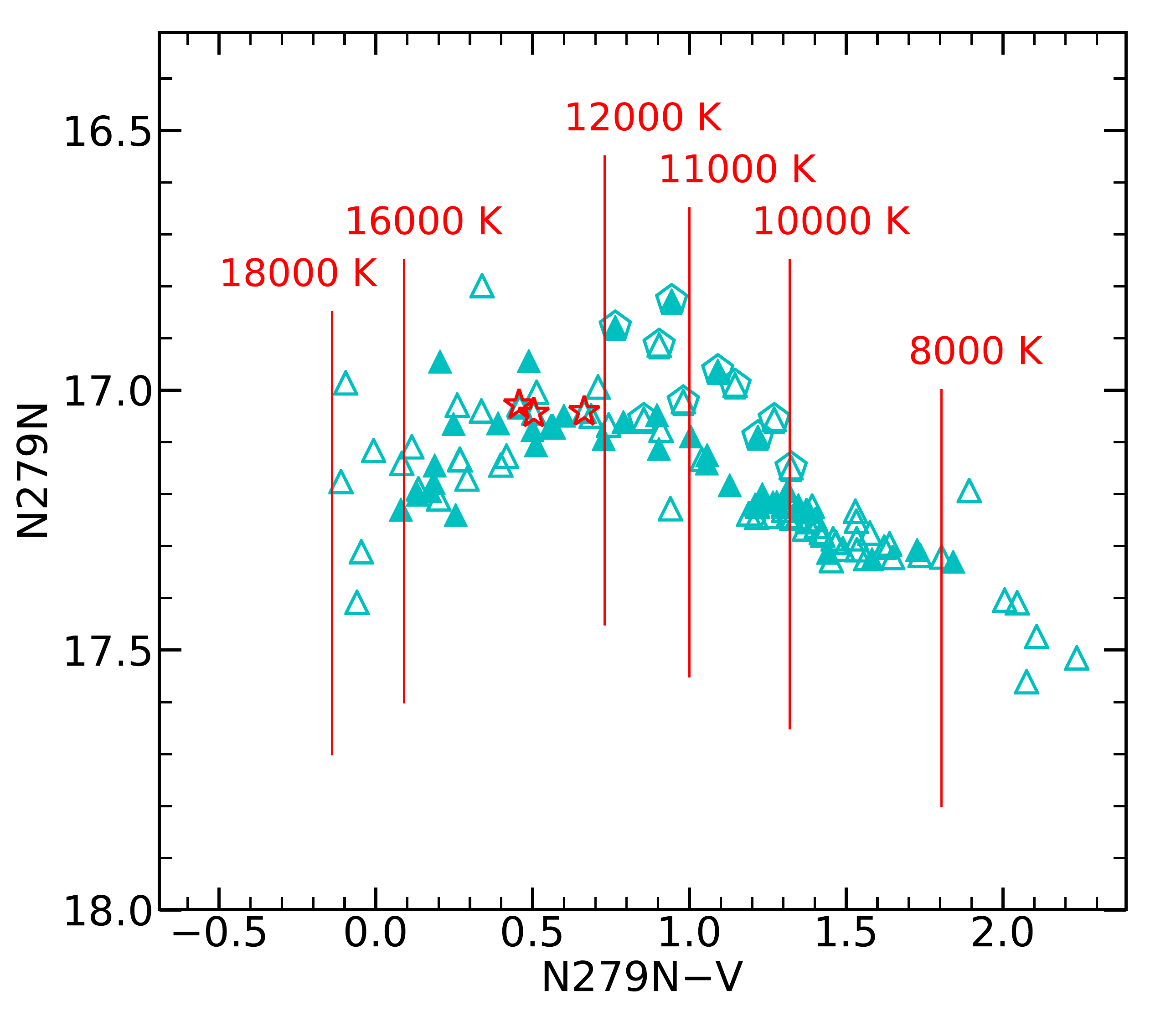}
\caption {A zoomed in view of the F148W$-$V vs F148W (upper panel) and N279N$-$V vs N279N CMDs (lower panel) showing only the BHB stars where the temperature locations for a log(g) = 3.5 and [Fe/H] = $-$1.5 are marked in red. The BHB gaps are marked with black arrows in the upper panel of the figure. The cyan pentagons are the over luminous BHB stars as classified by \citep{Moehler2014} in both the panels. The three red stars marked in the figures are B22, B186 and B302 \citep{Khalack2010}.}
\label{fuv_nuv_cmd_temp}
\end{center}
\end{figure}

\begin{table}
\centering
\caption{Temperature and reddening corrected colour of the BHB Gaps}
\resizebox{60mm}{!}{

\begin{tabular}{ccc}
\hline
\hline
Gap & F148W$-$V & $T_{eff}$ [K] \\\hline
G0 & 1.87 & 10,200\\
G1 & 1.48 & 11,000\\
G2 & 1.18 & 11,700\\
G3 & 0.97 & 12,300\\\hline
\label{hb_gaps}
\end{tabular}}
\end{table}

With the temperature estimations of the BHBs obtained from Section \ref{temp_color}, we attempt to identify possible gaps in the BHB distribution and their corresponding temperatures. In Figure \ref{fuv_nuv_cmd_temp}, we have shown F148W$-$V vs F148W and N279N$-$V vs N279N CMDs with only the BHB stars. We have marked the locations of the temperatures estimated from the respective colour - $T_{eff}$ relation. The gaps known in the literature are shown in black arrows in F148W$-$V vs F148W CMD (upper panel). In N279N$-$V vs N279N CMD of Figure \ref{fuv_nuv_cmd_temp} (lower panel), we notice that the distribution of BHB stars hotter than $T_{eff} \sim$ 10,500 K show a larger scatter in N279N magnitude. The separation between the two BHB groups around the Grundahl-jump (G-jump) \citep{Grundahl1999} are seen clearly in the FUV CMD as compared to the NUV CMD. The BHB stars hotter than G-jump show a flat distribution in FUV vs V CMD whereas it shows a curved distribution in NUV vs V CMD. The three red stars marked in both the panels in Figure \ref{fuv_nuv_cmd_temp} are B22, B186 and B302 which are spectroscopically confirmed to show a vertical stratification of iron in their atmospheres \citep{Khalack2010}. These stars are located in the FUV plateau near the G3 gap in F148W$-$V vs F148W CMD. The pentagons marked in the Figures are the 10 over-luminous BHB stars as classified by \cite{Moehler2014}. We note that the over-luminous HB stars show a jump in magnitude from that of BHBs which is more prominent in the $T_{eff}$ ranging from $\sim$ 10,200 K to $\sim$ 12,000 K in both the CMDs. 

We find gaps at $T_{eff} \sim$ 10,200 K, 11,000 K and 12,300 K in F148W$-$V vs F148W CMD which corresponds to the G0, G1 and G2 gaps respectively identified by \cite{Ferraro1998} in the HST FUV CMD of cluster M80. The gaps with F148W$-$V colour and the corresponding temperatures are given in Table \ref{hb_gaps}. In F148W$-$V vs F148W CMD, we identify the G-jump gap (G2) at $T_{eff} \sim$ 11,700 K which is close to the value identified by \cite{Grundahl1999} for this cluster. 

\subsection{Discussion}\label{discus_hb}
The BaSTI isochrone used for the comparison is able to reproduce the BHB distribution in UV CMDs upto T$_{eff}$ $\sim$ 11,500 K. Spectroscopic observations of the BHB stars by \cite{Moehler2014} revealed that the BHB stars hotter than $T_{eff} \sim$ 11,500 K suffer from the effects of atomic diffusion as a result of which they show an overabundance of metals. They found that the hotter BHB stars have lower He I abundances whereas Si II and Fe II abundances are higher as compared to the BHB stars with $T_{eff} <$ 11,500 K. As the effects of diffusion in BHB stars are not addressed in the BaSTI isochrones generated from the FSPS model, we see a significant deviation in the FUV luminosity between the model and the observed BHB distribution (Figures \ref{caf2_v_cmd}-\ref{nuv_cmd}) for stars with $T_{eff} >$ 11,500 K.  

The plateau in the FUV magnitudes found in the FUV$-$V vs FUV CMDs (Figures \ref{caf2_v_cmd} and \ref{caf2_n2_cmd}), start at about 11,500 K, indicating that the luminosity plateau could be a result of diffusion. The reduction in the FUV luminosity can be caused by the increased absorption in the FUV due to enhanced atmospheric metallicity from atomic diffusion. This study thus demonstrates that the FUV$-$V vs FUV CMDs could be used as an ideal proxy to detect the presence of diffusion among the BHB stars.

We notice two groups in the temperature distribution (Figure \ref{hist_colour}) that are mainly due to the BHB stars with and without the effect of diffusion. In the study of HB population in NGC 2808, \cite{Dalessandro2011} found that the observed distribution of T$_{eff}$ has four groups which can be well reproduced by synthetic models by assuming different initial He abundances with Y varying from 0.248 to 0.30. They concluded that the separate groups in the HB temperature distribution are due to the multiple stellar populations present in the cluster. \cite{Piotto2013} studied the multiple stellar phenomena in NGC 288 and found that the initial He difference in the two populations of MS and RGB is very small ($\Delta$Y $\sim$ 0.013). The Y abundance in the Kurucz and BaSTI models that is used for HB analysis in our study is $\sim$ 0.248. Thus, the two peaks seen in the temperature distribution of BHBs may not be arising due to the small variations in the initial He abundances. It is possible that these are actually caused due to the presence of gaps in the BHB distribution. The effects of atomic diffusion combined with variations in the initial He abundances, if any, need to be properly incorporated in the models in order to explain the UV properties of the BHB stars.

The NUV$-$V distribution of the BHB shows a curved profile, with bluer and the redder BHB stars showing fainter NUV magnitudes. A comparison of bottom panel of Figure \ref{fuv_nuv_cmd_temp} with the Figure 8 of \cite{Moehler2014} indicates that the NUV magnitude profile directly correlates with the Mg II strength. The cooler (T$_{eff} <$ 11,000 K) and the hotter stars (T$_{eff} >$ 14,000 K) have relatively large Mg II strength, resulting in fainter N279N magnitude. Therefore, the NUV magnitudes presented here closely match with the estimated Mg II abundances.

According to the study by \cite{Momany2004}, the HB of NGC 288 terminates at 16,000 K whereas our analysis suggests that the HB population extends till 18,000 K (Figure \ref{hist_colour}). This is primarily due to the contribution of stars located outside the HST FOV, which in turn shows the advantage of combining UVIT with the ground and GAIA data owing to their larger FOV. 
 
\section{SEDs of bright gap object and EHB stars}\label{temp_ehb}
We have derived the parameters (Luminosity, temperature and radius) of the two candidate EHB stars and the bright gap object using virtual observatory tool, VOSA (VO SED Analyser, \cite{Bayo2008}). Python SED fitter program is useful for generating the SEDs of a large number of samples at one go whereas it is not possible in VOSA. On the other hand, VOSA is useful for an in depth analysis of the SEDs where one can manually change or fix the model free parameters and check for any UV or IR excess. VOSA calculates the synthetic flux for a selected theoretical model using the filter transmission curves. The synthetic fluxes are then scaled with the observed fluxes by fixing the distance of the object. It does $\chi^{2}$ minimisation test to find the best fit parameters of the SED. We have used Kurucz models \citep{Castelli1997} for fitting the SEDs by adopting a distance of 8.8 Kpc, fixing the metallicity close to that of the cluster i.e. [Fe/H] = $-$1.5, and taking log(g) = 5.0 by assuming them to be sub-dwarfs. The radius of the objects were calculated from the scaling factor which is equal to $\big(R/D)^{2}$, where R is the radius and D is the distance to the object. 

\begin{figure}
\begin{center}
\includegraphics[scale=0.3]{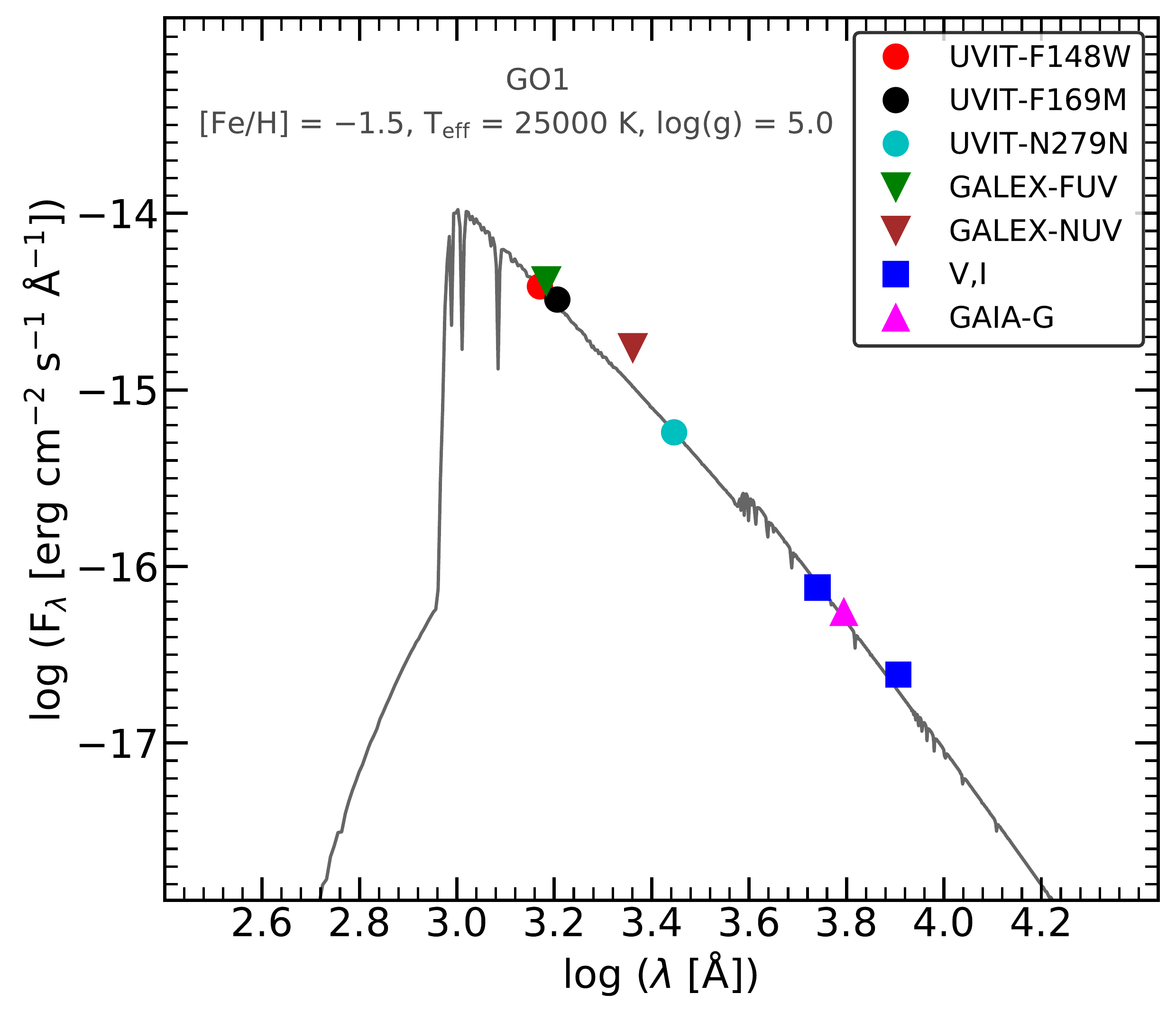}\\
\includegraphics[scale=0.3]{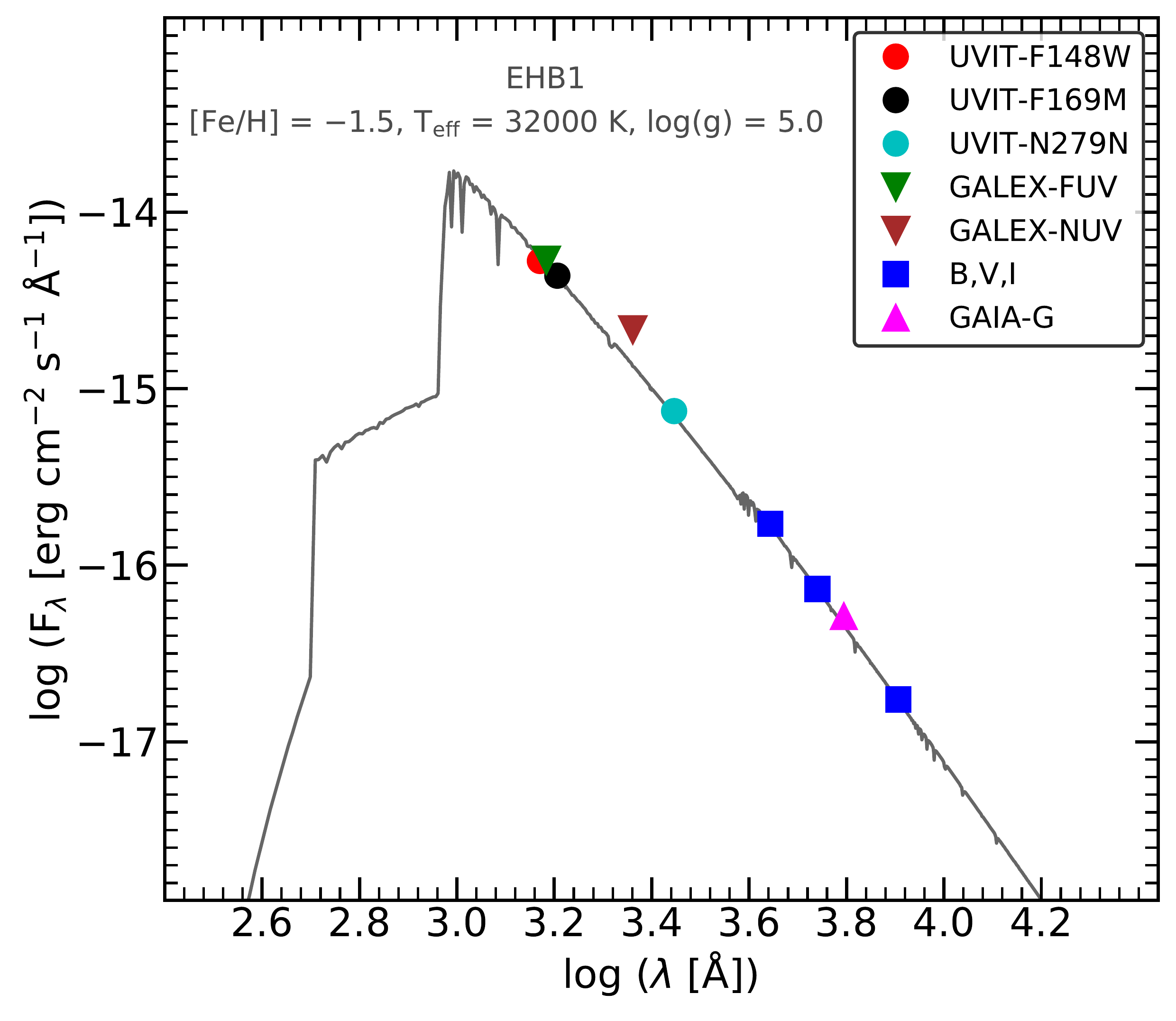}\\
\includegraphics[scale=0.3]{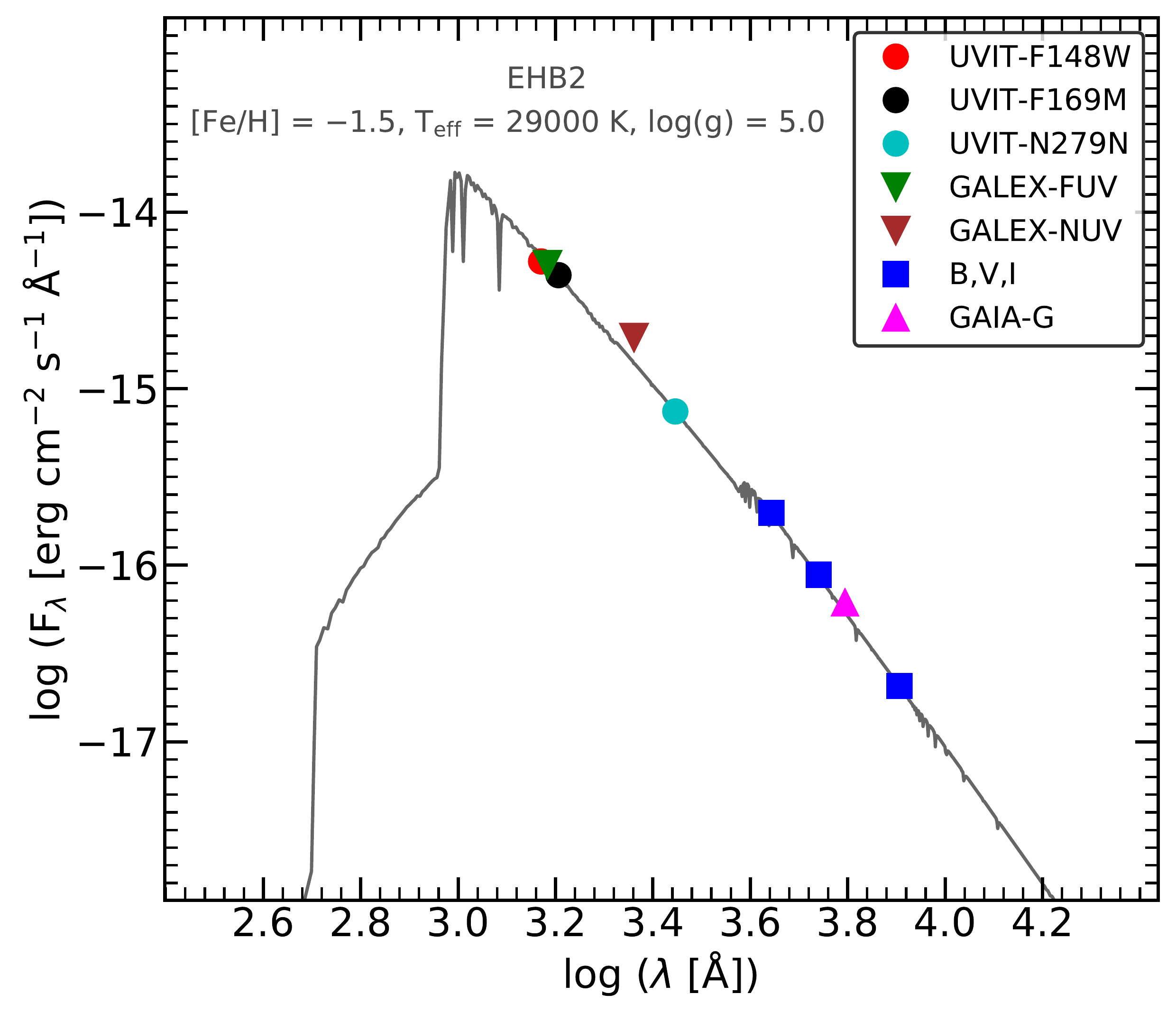}
\caption{Spectral energy distribution (SED) of the bright gap object and EHB stars after correcting for extinction. The best fit parameters are mentioned in the figure.}
\label{sed_ehb}
\end{center}
\end{figure}

We have generated the SEDs of two candidate EHB stars designated as EHB1 and EHB2 by combining the flux measurements of UVIT (F148W, F169M, N279N) with GALEX (FUV, NUV), GAIA (G) \citep{Gaia2016,Gaia2018} and Ground (B, V, I). For the bright gap object, we have generated the SED by combining the flux measurements of UVIT (F148W, F169M, N279N) with GALEX (FUV, NUV), GAIA (G) \citep{Gaia2016, Gaia2018} and HST (V, I) pass bands. The SEDs of the 3 stars are shown in Figure \ref{sed_ehb}. We have obtained the GALEX FUV and NUV band fluxes of these objects by running PSF photometry \citep{Stetson1987} on the FUV and NUV intensity maps of the cluster (Tile Name: MIS2DFSGP\_30531\_0144). The best fit SED parameters of the gap object and EHB stars are given in Table \ref{sed_par_ehb}. We find that the UVIT FUV flux measurements are consistent with GALEX FUV measurements whereas the GALEX NUV flux measurements show deviations from the model expected fluxes. These stars are located in the crowded regions close to the cluster centre where the GALEX NUV band suffers from poor resolution. This could lead to an overestimation of the NUV flux due to the contamination from nearby stars. The reduced $\chi^{2}$ ($\chi^{2}_{reduced}$) value mentioned in the Table \ref{sed_par_ehb} are obtained after excluding the GALEX NUV data point.
	
From Table \ref{sed_par_ehb}, we notice that the EHB stars have similar luminosity whereas the bright gap object is $\sim$ 1.3 times less luminous than the EHB stars. All of them have similar radii (R $\sim$ 0.2 R$_{\odot}$). The temperatures of the EHB1 and EHB2 ($T_{eff}\sim$ 32,000 and 29,000 K respectively) clearly suggest that they belong to the class of EHBs/subdwarfs \citep{Heber1986}. The temperature and radius of the bright gap object along with its location in the UV CMDs suggest that it is likely to be a subdwarf. The large $\chi^{2}_{reduced}$ value (46.38) for the SED fit (even after excluding the GALEX NUV data point), as compared to the two EHB stars, is due to the excess observed flux in the I band, excluding which reduces the value to 5.92. This might suggest the  presence of a cooler companion which can be verified if we include the observations from the longer wavelengths in the SED. The EHB stars are located at around the half-light radius of the cluster ($r_{h} \sim 2\farcm23$) whereas the bright gap object is located near the core radius of the cluster ($r_{c} \sim 1\farcm35$, \cite{mc2005}).

\begin{table*}
\centering
\caption{SED fit parameters of bright gap object and EHB candidates}
\resizebox{130mm}{!}{

\begin{tabular}{ccccccc}
\hline
\hline
Star ID & [Fe/H] & T$_{eff}$ [K] & log(g) & L/L$_{\odot}$ & R/R$_{\odot}$ &  $\chi^{2}_{reduced}$  \\\hline
GO1 & -1.5 & 25000$\pm$500 & 5.0$\pm$0.25 & 16.60$\pm$0.04& 0.22$\pm$0.01 & 46.38\\\hline
EHB1 & -1.5 & 32000$\pm$500 & 5.0$\pm$0.25 & 21.93$\pm$0.05 & 0.15$\pm$0.01 & 14.66 \\
EHB2 & -1.5 & 29000$\pm$500 & 5.0$\pm$0.25 & 22.39$\pm$0.05 & 0.19$\pm$0.01 & 10.56 \\\hline
\label{sed_par_ehb}
\end{tabular}
}
\end{table*}

\subsection{Discussion}
Generally, we see a large number of EHB population with a well defined sequence in the CMDs of massive GCs such as NGC 2808, $\omega$Cen, M 54, NGC 6752 \citep{Momany2004, Dalessandro2011} whereas in low mass GCs they are less in number and are hard to detect at the faint HB end of the optical CMDs. In these cases, UV CMDs are very useful where the EHB stars are very bright and form a separate sequence as compared to the optical CMDs. \cite{Momany2004} reported a discontinuity/gap at $T_{eff} \sim$ 23,000 K present in the massive GCs with EHB stars and suggested that these are early Helium flashers. Two such EHBs/subdwarfs detected in the cluster NGC 288 have temperatures $>$ 23,000 K consistent with the previous studies \citep{Momany2004}. The SED analysis of bright gap object suggests that this could be a subdwarf-binary candidate with a cooler companion. This is also supported by the fact that most of the binary systems are present within 1$r_{h}$ of the cluster where the binary fraction $8\% < f_{b} < 38 \%$ is very high as compared to the cluster outskirts \citep{Bellazzini2001}. This coincides with the positions of EHB stars within 1$r_{h}$ of the cluster suggesting that the binary scenario could be the dominant formation channel for the three subdwarfs in the cluster \citep{Podsiadlowski2008}.

\section{Properties of BSS\lowercase{s}}\label{param_bss}
\subsection{Parameters of BSSs from Photometry}
\begin{figure}
\begin{center}
\includegraphics[scale=0.33]{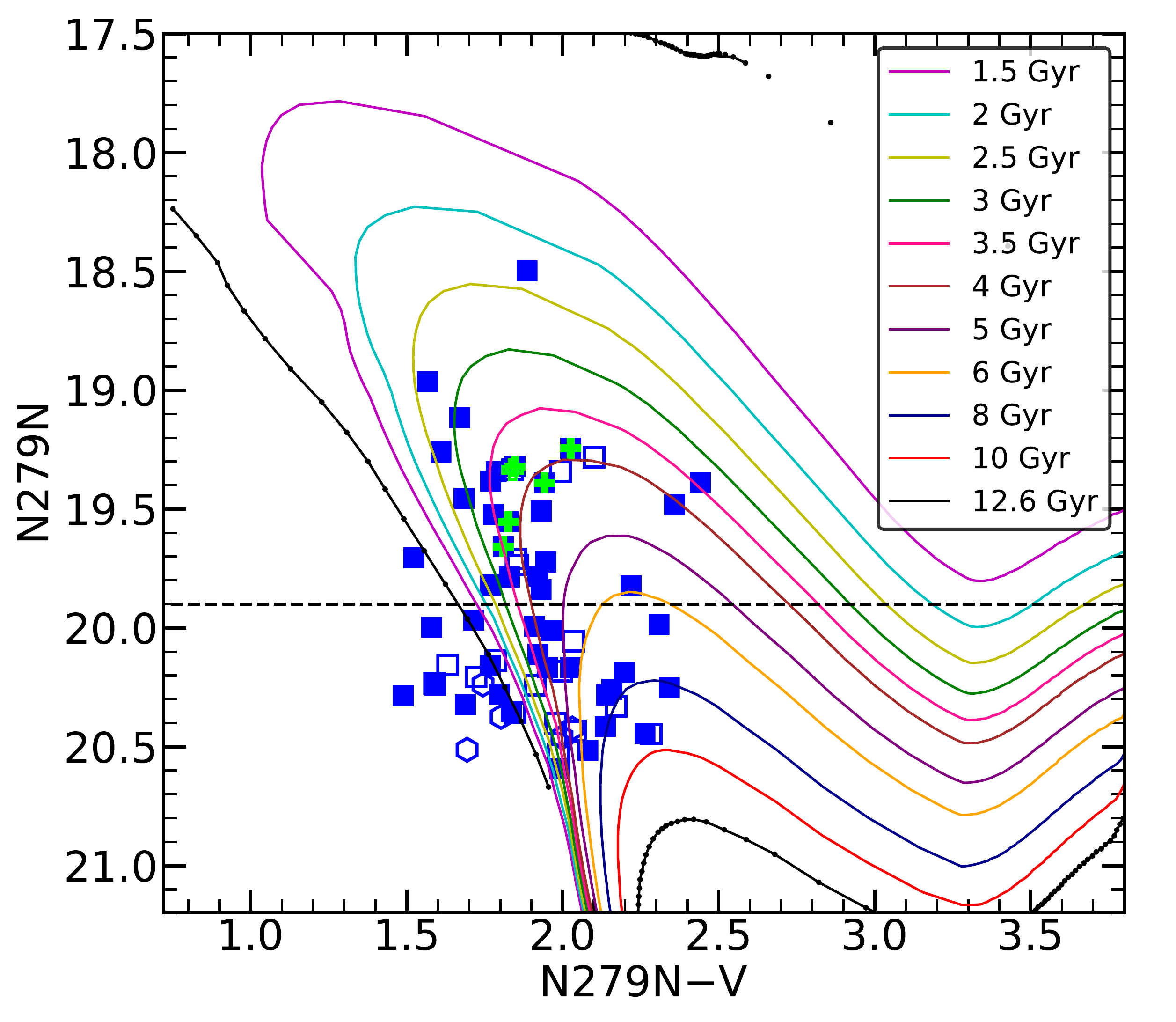}
\caption{A zoomed in view of Figure \ref{nuv_cmd} where only identified BSSs are shown in the N279N$-$V vs N279N CMD. BaSTI isochrones of [Fe/H] = $-$1.28, Y = 0.248 and different ages marked in the legend are over plotted on the CMD. The black dashed line divides the BSSs sample into bright and faint BSSs.} 
%Lower panel : KDE plot of the BSSs distribution. 
\label{bss_prop}
\end{center}
\end{figure}

In order to estimate the parameters of 68 BSS members selected from the N297N$-$V vs N279N CMD and GAIA DR2 data, we have used BaSTI isochrones of ages ranging from 1.5 Gyr to 10 Gyr with [Fe/H] = $-$1.28 and Y = 0.248 in such a way that it covers the location of BSS region in the CMD. These isochrones, over-plotted in the observed CMD are shown in Figure \ref{bss_prop}. We have divided the BSS sample into two groups based on the N279N magnitude shown as black dashed line in the Figure. In our sample, the BSSs with N279N $<$ 19.95 are the Bright BSSs (BBSSs) whereas those with N279N $>$ 19.95 are the Faint BSSs (FBSSs). The BBSSs are less in number and span a larger range in magnitude as compared to the FBSSs which have a broader colour distribution.

\begin{figure}
\begin{center}
\includegraphics[scale=0.39]{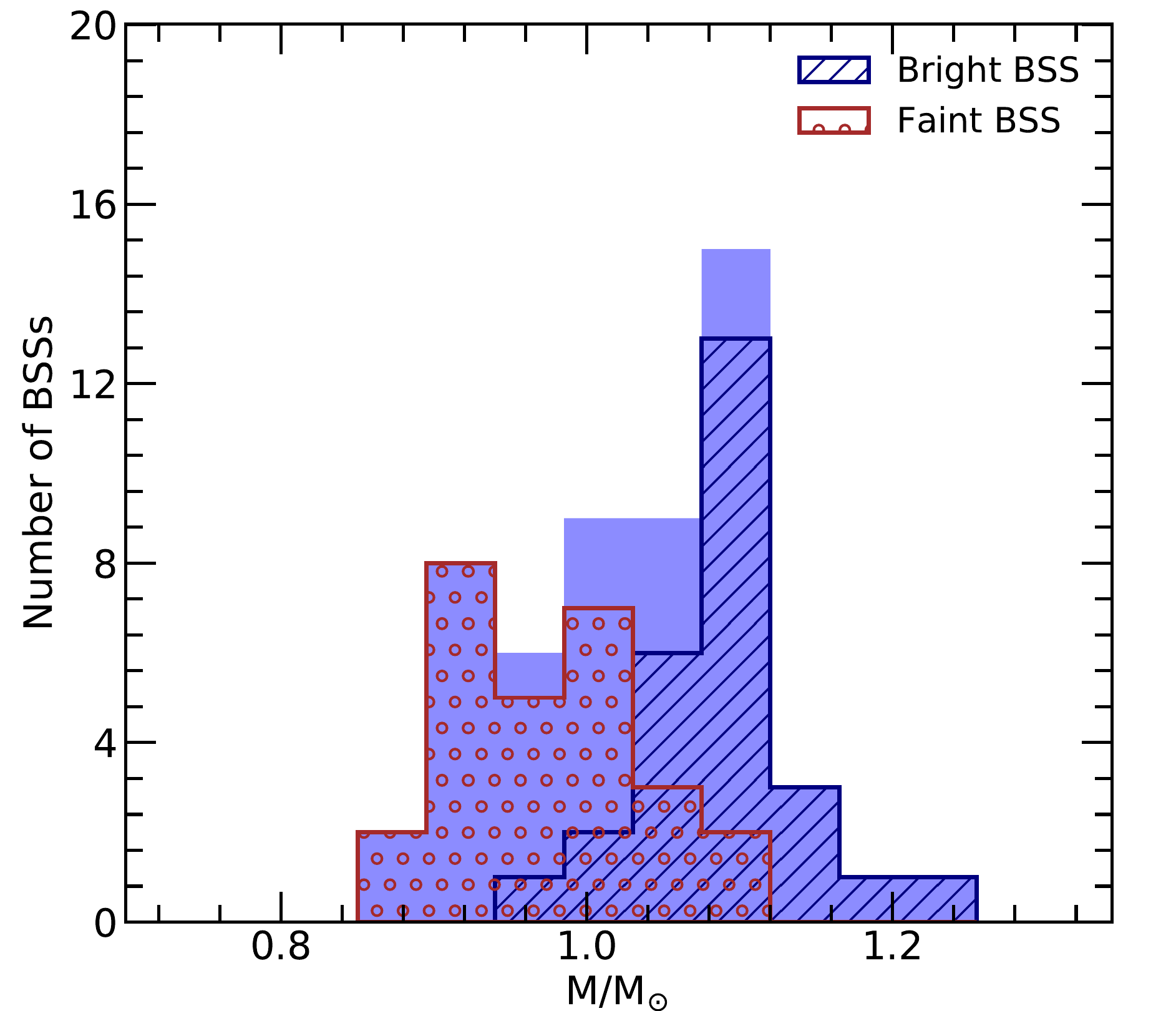}
\includegraphics[scale=0.395]{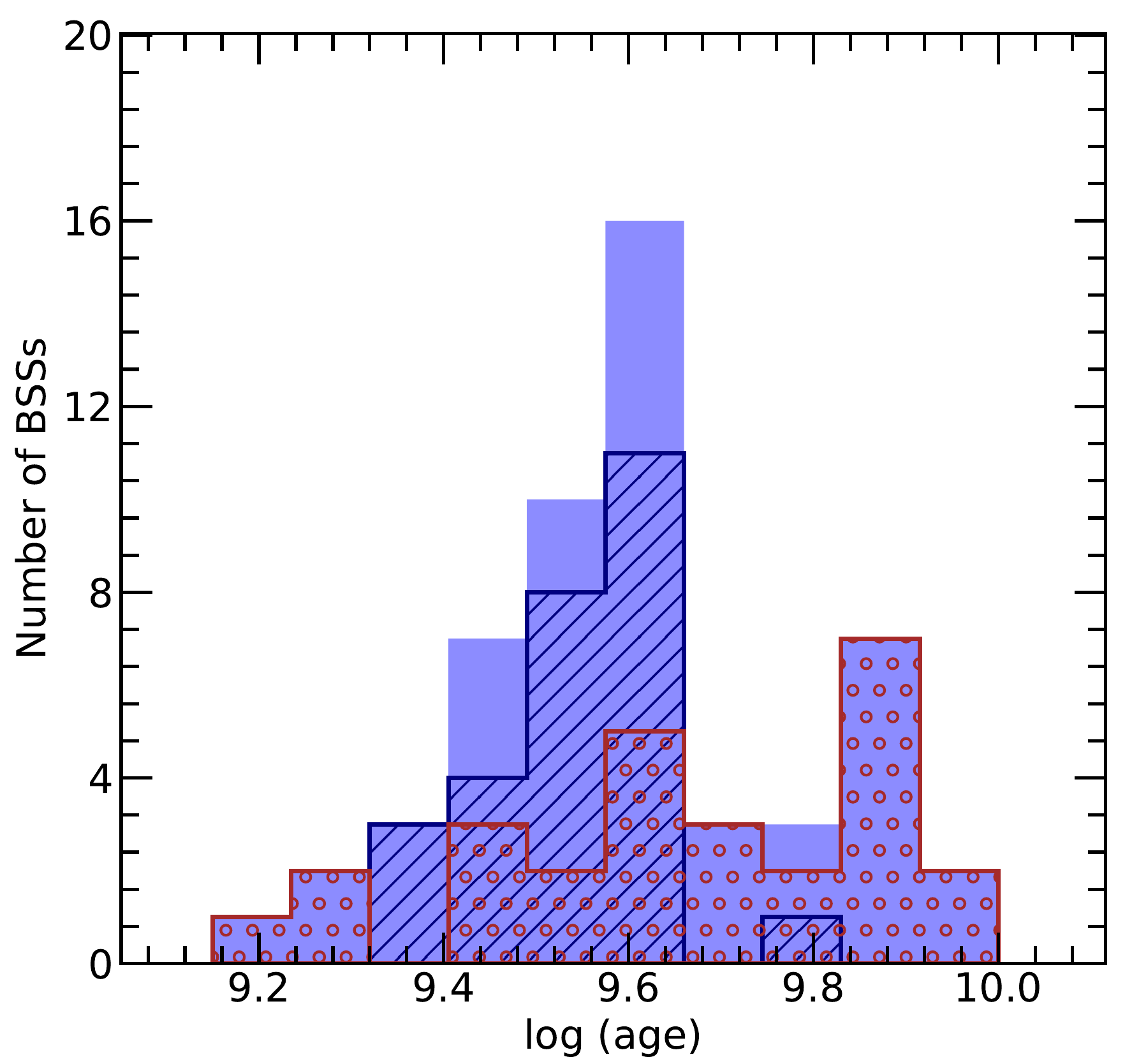}
\caption{Mass (upper panel) and age (lower panel) distribution of 54 BSSs (light blue shaded), 22 BBSSs (blue hatched) and 32 FBSSs (brown circles).}
\label{bss_hist}
\end{center}
\end{figure}

We have derived the parameters of the BSSs by cross-matching the N279N$-$V colour and N279N magnitude with those of the model colour and magnitude of BaSTI isochrones of different ages. The histogram of mass and age of 54 BSSs (22 BBSSs, 32 FBSSs) are shown in Figure \ref{bss_hist}. From the upper panel of the figure, we notice that the mass of the BSSs range from 0.86 M$_{\odot}$ to 1.25 M$_{\odot}$ with a peak at $\sim$ 1.03 M$_{\odot}$. The mass of the FBSSs range from 0.86 - 1.1 M$_{\odot}$ with an average mass $\sim$ 0.97 M$_{\odot}$ whereas those of the BBSSs range from 0.98 - 1.25 M$_{\odot}$ with an average mass $\sim$ 1.1 M$_{\odot}$, which are less than 2M$_{TO}$ ($\sim$ 1.56 M$_{\odot}$) of the cluster.

The ages of the BSSs range from 2.0 - 10 Gyr with a peak at $\sim$ 4 Gyr as shown in the lower panel of the Figure \ref{bss_hist} which are in agreement with \cite{Bellazzini2001}. The BBSSs are younger having an average age of $\sim$ 3.5 Gyr whereas the FBSSs have a large range in age, with an average age of $\sim$ 4.5 Gyr. The brightest BSS is of age $\sim$ 2.5 Gyr and mass $\sim$ 1.25 M$_{\odot}$. This BSS is located at a radius of $\sim$ 0$\farcm$66 from the cluster centre. There are $\sim$ 2 BSSs in N279N$-$V vs N279N CMD shown in Figure \ref{bss_prop} which are redder than the other BSSs in the CMD. These BSSs could have evolved from the MS and are currently in the Sub-giant phase and potential candidates for the Yellow Straggler Stars (YSSs). The masses of the SX Phe variables are $\sim$ 1.07 M$_{\odot}$ which are in agreement with the masses of SX Phe variables estimated by \cite{Fiorentino2015} for different clusters.

\subsection{SEDs of FUV detected BSSs}
We have generated the SEDs of the BSSs detected in FUV using VOSA which we also used for the SEDs of EHB stars as described in Section \ref{temp_ehb}. We have constructed the SEDs of the BSSs by combining the UVIT fluxes (3 filters) with the optical fluxes (V, I), thus, covering the spectral range from optical to UV wavelengths. We have considered the same cluster parameters for the SED fitting as mentioned earlier in Section \ref{temp_ehb}. The SED fit parameters (T$_{eff}$, luminosity and radius) of the BSSs along with the errors are given in Table \ref{sed_par_bss} where the ages and masses are derived from the BaSTI isochrones. The errors are small as these are arising only from the SED fits. All the FUV detected BSSs fall under BBSSs category except three of them which lie in the FBSSs category. Out of 15 BSSs, we found 11 BSSs (9 BBSSs and 2 FBSSs) with $\chi^{2}_{reduced} > 20$. The deviations of the observed fluxes from the model are due to FUV excess which are significant beyond the 3$\sigma$ predictions of the model. This points to the possibility of the presence of a hot companion such as WD associated with the BSSs. These stars will be studied in detail in future.  
  
\begin{figure}
\begin{center}
\includegraphics[scale=0.28]{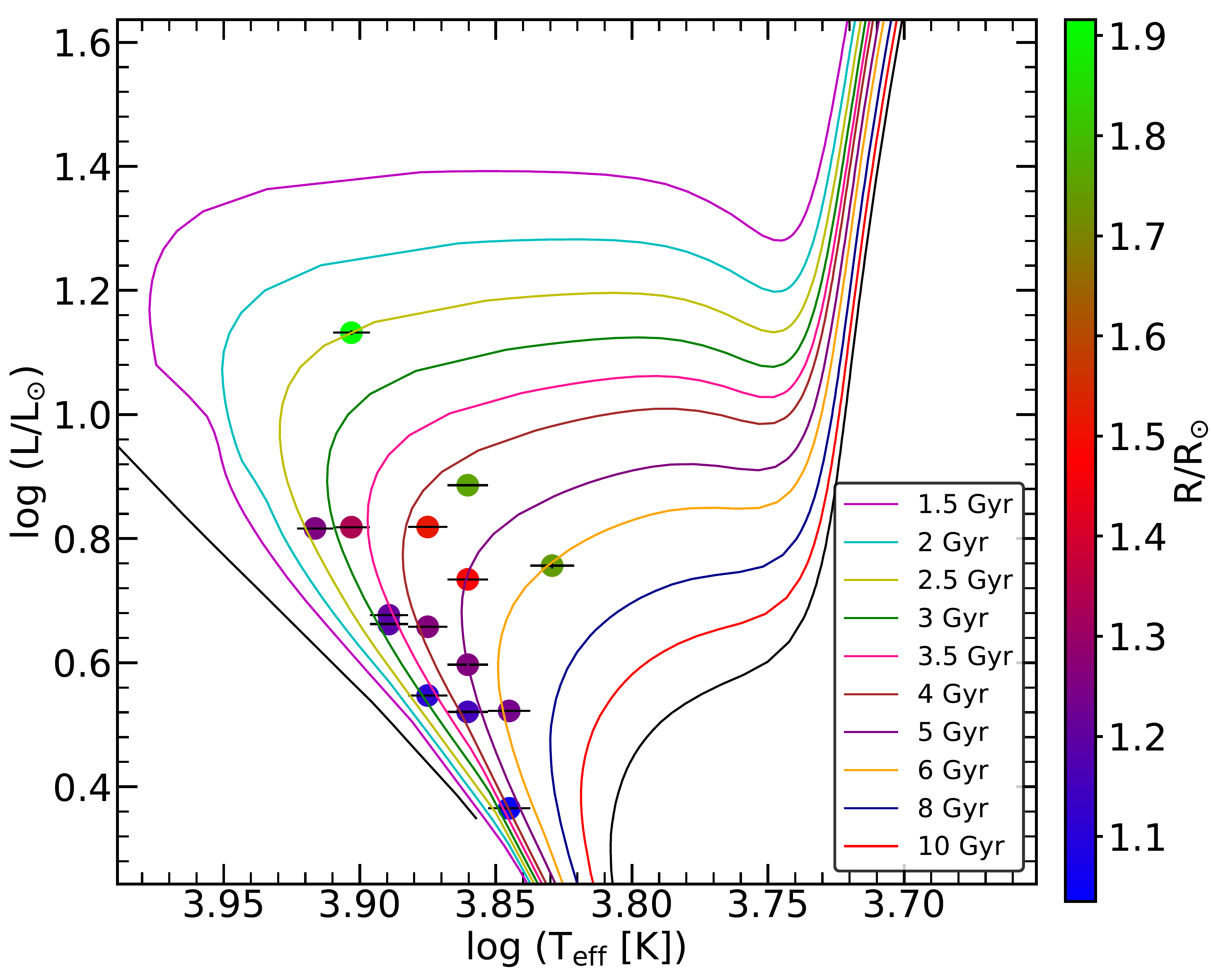}
\caption{FUV detected BSSs plotted in the log L vs log $T_{eff}$ H-R diagram based on the parameters derived from SED fitting where the colour bar represents the estimated radii. The parameters of the BaSTI isochrones plotted are similar to Figure \ref{bss_prop}.}
\label{bss_hr}
\end{center}
\end{figure}

For a better understanding of the present evolutionary phases of the FUV detected BSSs, we have plotted them in the H-R diagram as shown in Figure \ref{bss_hr}. The colour bar shows the radii of the BSSs estimated from the SED fitting. The BaSTI isochrones plotted are similar to the Figure \ref{bss_prop}. Here, we notice that the age of the BSSs peaks at $\sim$ 4 Gyr which is in agreement with the photometric estimates. The brightest BSS in the H-R diagram is of age 2.5 Gyr with a larger radius of R $\sim$ 1.9 R$_{\odot}$ as compared to other BSSs. 

We have also constructed the SED of the Evolved BSS candidate (Section \ref{bss_phot}) by combining UVIT flux (N279N) with the HST (V, I) and GAIA (G) fluxes. The temperature ($T_{eff} \sim$ 5500 K) and luminosity of the object (L$\sim$ 66.59$\pm$1.5 L$_{\odot}$) derived from the SED fit are in close agreement with \cite{ferraro2016}. These estimations along with the proper motion membership from the GAIA DR2 clearly suggests that it can be very likely a EBSS candidate. This serves as a good spectroscopic target for future study of the evolution of BSSs.  

\begin{table*}
\centering
\caption{SED fit parameters of FUV detected BSSs. Columns 2, 3 and 4 present the temperatures, luminosities and radii of BSSs derived from the SED fitting whereas columns 5 and 6 present the masses and ages of BSSs derived from the BaSTI isochrones.}
\resizebox{120mm}{!}{
\begin{tabular}{ccccccc}
\hline
\hline
Star ID & $T_{eff}$ [K] & $L/L_{\odot}$ & $R/R{_\odot}$ & $M/M{_\odot}$ & Age [Gyr] \\\hline
BSS1	&	8000	$\pm$ 125	&	13.55	$\pm$	0.12	&	1.92	$\pm$	0.06	&	1.25	&	2.51	\\
BSS2	&	7750	$\pm$ 125	&	4.75	$\pm$	0.02	&	1.21	$\pm$	0.04	&	1.07	&	3.47	\\
BSS3	&	7250	$\pm$ 125	&	5.42	$\pm$	0.01	&	1.48	$\pm$	0.05	&	1.01	&	5.01	\\
BSS4	&	6750	$\pm$ 125	&	5.71	$\pm$	0.06	&	1.75	$\pm$	0.07	&	0.98	&	6.03	\\
BSS5	&	7500	$\pm$ 125	&	4.55	$\pm$	0.01	&	1.26	$\pm$	0.04	&	1.02	&	4.47	\\
BSS6	&	7250	$\pm$ 125	&	3.95	$\pm$	0.01	&	1.26	$\pm$	0.04	&	0.99	&	5.01	\\
BSS7	&	7500	$\pm$ 125	&	3.53	$\pm$	0.01	&	1.11	$\pm$	0.04	&	1.01	&	3.98	\\
BSS8	&	7000	$\pm$ 125	&	3.33	$\pm$	0.01	&	1.24	$\pm$	0.04	&	0.94	&	6.03	\\
BSS9	&	8250	$\pm$ 125	&	6.55	$\pm$	0.01	&	1.25	$\pm$	0.04	&	1.17	&	2.51	\\
BSS10	&	8000	$\pm$ 125	&	6.58	$\pm$	0.01	&	1.34	$\pm$	0.04	&	1.14	&	3.02	\\
BSS11	&	7000	$\pm$ 125	&	2.32	$\pm$	0.01	&	1.04	$\pm$	0.04	&	0.96	&	3.98	\\
BSS12	&	7250	$\pm$ 125	&	3.32	$\pm$	0.01	&	1.15	$\pm$	0.04	&	1.00	&	3.98	\\
BSS13	&	7750	$\pm$ 125	&	4.60	$\pm$	0.01	&	1.19	$\pm$	0.04	&	1.09	&	3.02	\\
BSS14	&	7500	$\pm$ 125	&	6.59	$\pm$	0.01	&	1.52	$\pm$	0.05	&	1.08	&	3.98	\\
BSS15	&	7250	$\pm$ 125	&	4.60	$\pm$	0.01	&	1.76	$\pm$	0.06	&	1.06	&	4.46	\\
\hline
\label{sed_par_bss}
\end{tabular}
}
\end{table*}

\subsection{Radial Distribution}
BSSs being more massive than the HB and RGB stars, are expected to segregate towards the cluster centre due to dynamical friction, on the timescale of the order of half-mass relaxation time ($t_{r,h}$), which is $\sim$ 2 Gyr for this cluster \citep{harris1996, harris2010}. Since $t_{r,h}$  timescale is much shorter than the age of the cluster, the mass segregation should have taken place long ago with the BSSs being more centrally concentrated than the HB and RGB stars. In order to check this, we have derived the radial distribution of the BSSs and HBs. For transforming the RA and DEC coordinates of these stars to XY system, we have used the cluster centre at RA = $00^{h}52^{m}45.24^{s}$, DEC = $-26^{\circ}34'57.4''$ given by \cite{Goldsbury2010} and estimated their radial distance from it. For constructing the radial distribution, we have sampled the stellar populations up to a radius of 7$'$ from the cluster centre. The final sample used for the radial distribution consists of 68 BSSs (28 BBSSs, 40 FBSSs) and 115 BHB stars.

\begin{figure}
\begin{center}
\includegraphics[scale=0.3]{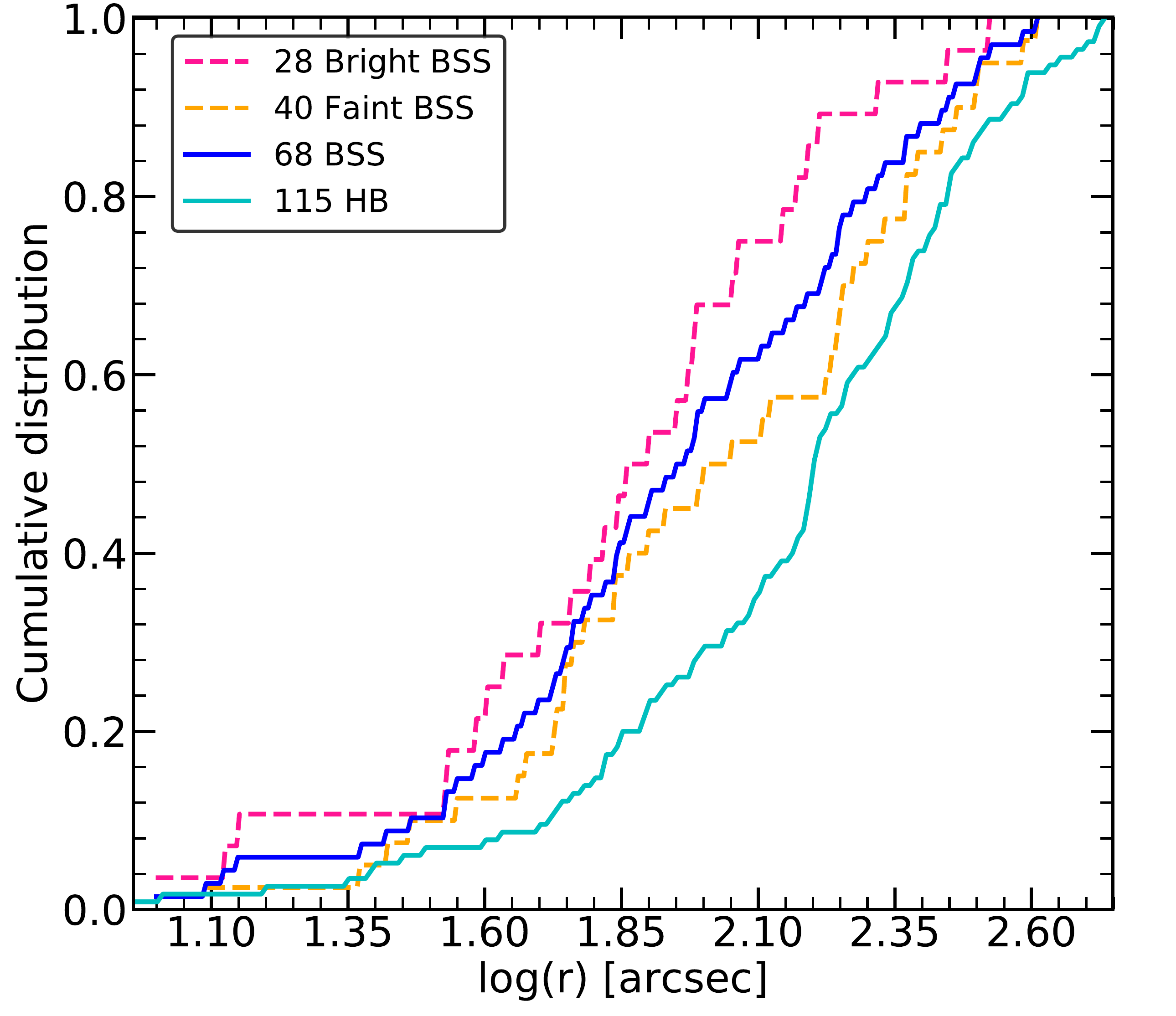}
\caption{Cumulative radial distribution of BBSS, FBSS and BSS population with respect to the HB population.}
\label{bss_cum}
\end{center}
\end{figure}

The cumulative radial distribution of the BBSSs, FBSSs, BSSs and BHB populations are shown in the Figure \ref{bss_cum}. We have considered the BHB population as the reference population and performed K-S test to check the statistical significance of the differences among the radial distribution of the above stellar populations. According to the K-S test, the distribution of BSS population is significantly different from the BHB population with a p-value of $\sim$ 3$\times$10$^{-4}$. This shows that the BSS population is centrally concentrated than the BHB population. The K-S test also suggests that the BBSSs are more centrally concentrated than the FBSSs with p-values of $\sim$ 8$\times$ 10$^{-4}$ and $\sim$ 0.12 respectively with respect to the reference BHB population. The probability that the BBSS and FBSS distributions are drawn from the same population is only 5.6$\%$. The central concentration of BBSSs is likely due to their relatively large mass as compared to the FBSSs.

\subsection{Specific frequency}
\begin{figure}
\begin{center}
\includegraphics[scale=0.34]{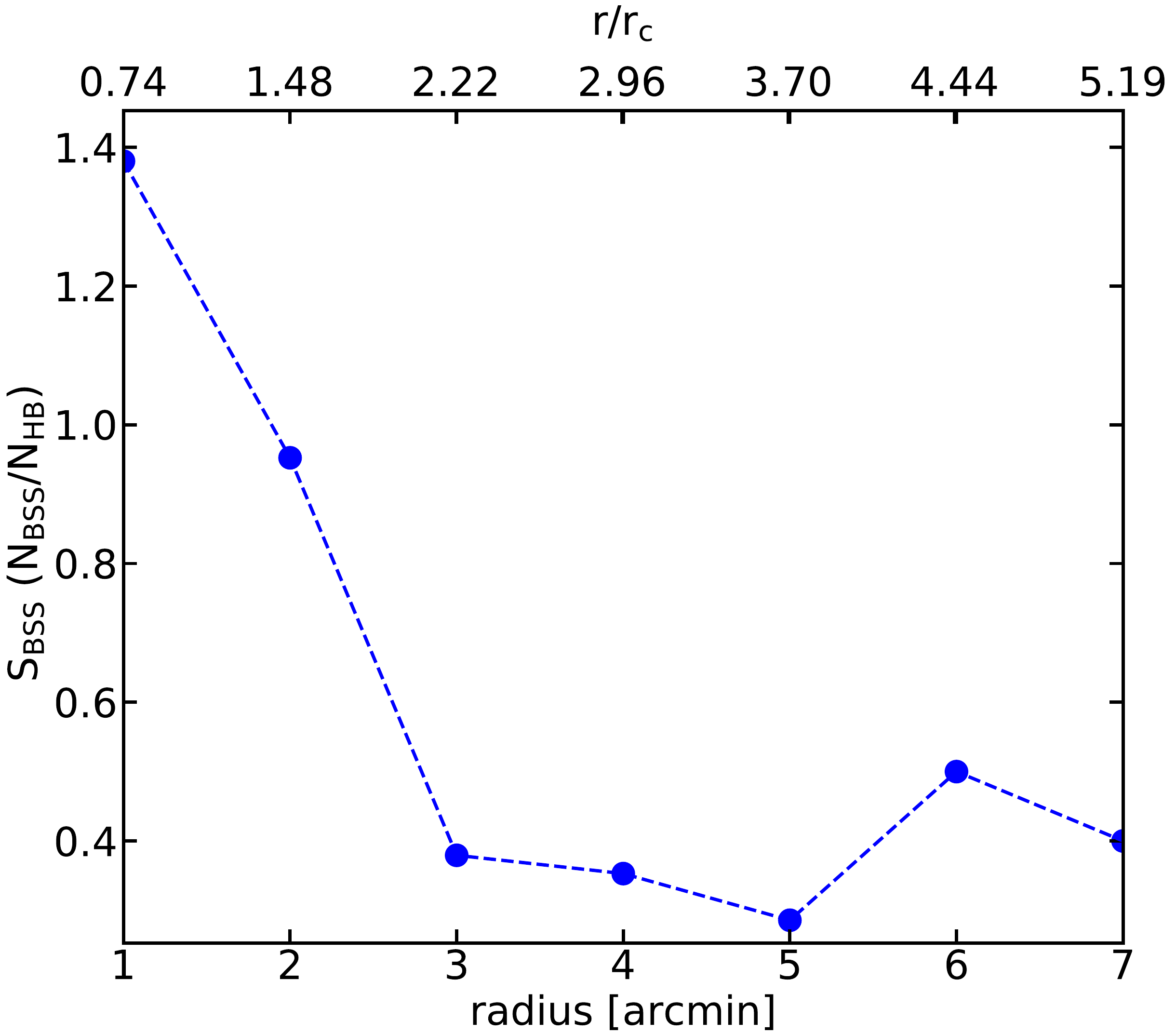}\\
\caption{Specific frequency of BSSs with respect to HB stars plotted as a function of radial distance and r/r$_{c}$ from the cluster centre.}
\label{sbss}
\end{center}
\end{figure}

The specific frequency of BSSs ($S_{BSS}$) is defined as the ratio between the number of BSSs ($N_{BSS}$) and the number of HB stars ($N_{HB}$) which are observed in the same region. A radial profile of this parameter helps us to understand the distribution of BSSs with respect to BHB stars in a cluster. We have derived the specific frequency of BSSs ($S_{BSS}$) by dividing the observed cluster region into several concentric annuli (each of 1$'$) and then counting the number of BSSs and HB stars in each annulus. The $S_{BSS}$ as a function of the radial distance from the cluster centre (lower axis) and $r/r_{c}$ (upper axis) is shown in Figure \ref{sbss}. Our estimation of the $S_{BSS} = 1.38 \pm 0.45$ in a region $r < 1r_{c}$ is consistent with that estimated by \cite{Bellazzini2001}. Here, the error in the $S_{BSS}$ is calculated by adopting the relation given by \cite{Sabbi2004}. The number of BSSs within a radius $\sim 1'$ of the cluster is $\sim$ 22 which is close to the number ($\sim$ 24) determined by \cite{Ferraro2003} for the cluster. 

We notice that most of the BSSs are concentrated in the region $r < 2r_{c}$. In addition to the central peak, we also find a second peak $S_{BSS}$ at $r \approx 4.5r_{c}$ with a sudden drop at $r \sim 3r_{c}$. This region with a sudden drop in $S_{BSS}$ is defined as the zone of avoidance \citep{Mapelli2004} which is at $r_{min} = 4'$ for the cluster in accordance with \cite{Lanzoni2016}. This shows that the cluster has a bimodal BSS distribution. The second peak (r$\sim 6'$) in the $S_{BSS}$ is genuine as the BSSs in the cluster outskirts are proper motion members according to GAIA DR2 \citep{Gaia2018b}.

\subsection{Discussion}\label{discus_bss}
The observed BSS distribution in the NUV-V vs V CMD (Figure \ref{bss_prop}) closely matches with the distribution shown in Figure 8 by \cite{Ferraro2012} for NGC 288. The BSSs age peaks at $\sim$ 4 Gyr (ranging from 2-10 Gyr) with most of them having ages older than the half-mass relaxation time of the cluster ($t_{rh} \sim$ 2 Gyr \cite{harris1996, harris2010}). This is in agreement with the prediction of constant formation of BSS in the last 7 Gyr by \citep{Ferraro2003}. The brighter BSSs are found to be more massive and younger than the fainter ones. According to the two BSS evolutionary scenarios proposed by \citep{Bellazzini2001}, the evolutionary mass transfer scenario puts a lower mass limit of 0.92 M$_{\odot}$ for the BSS in the cluster which is more or less in agreement with our findings. 

From the cumulative radial distribution, we found the BBSSs to be centrally concentrated as compared to the FBSSs unlike the previous studies on massive GCs such as M80, M15 \citep{Dieball2010, Haurberg2010} where it was the other way around. The authors anticipated that the interactions of the bright BSSs with the cluster members in the high density environments such as central regions would have kicked them out of the cluster. Being a low density cluster, the dynamical interactions in NGC 288 are expected to be less active in the BSS formation as compared to the high density clusters ($\rho_{c}\sim 10^{5}$ $L_{\odot}/pc^{3}$ \cite{mc2005}). \citep{Bellazzini2001} suggested that the large population of BSSs found in this cluster may be a result of their formation via mass transfer/coalescence of primordial binary systems. They found a high binary fraction within the 1$r_{h}$ of the cluster where 65 $\%$ of our BSS sample resides in the cluster.
 
The two peaks found in the BSS specific frequency of the cluster suggests that the cluster is of intermediate dynamical age and falls in the Family II category as per the definition given by \cite{Ferraro2012}. This in turn, indicates that the dynamical friction has affected the BSS population only in the central regions of the cluster till the $r_{min} \sim 5'$ resulting in the mass segregation of the massive BSSs. The presence of a few BSSs in the cluster outskirts beyond 5$'$ also conveys that the dynamical friction has not yet affected the BSS population in the outer regions. 

\section{Conclusions}\label{conclude}
Until now, the UV studies of GCs are mainly done using the high spatial resolution images obtained from the HST. With the successful launch and operations of the ASTROSAT, the UVIT has started producing good quality images of GCs in the FUV and NUV bands. The large field of view combined with good image quality makes the UVIT studies of stellar populations in the GCs very significant. Here, we present the results from our study of NGC 288 imaged using the UVIT during the first year of its operations. As the UVIT images resolve the core of the cluster in the FUV and NUV, we have analysed the complete sample of BHB in FUV and BSS in NUV thus, covering the full cluster region ($\sim$ 10$'$) in UV and highlighting the importance of UV observations. This study also highlights the usefulness of combining UVIT data with HST, ground and GAIA data. Our UVIT imaging study of NGC 288 has led us to the following conclusions:
\begin{enumerate}
    \item The UV bright stars in this cluster consist of 115 BHBs, 2 RRLs, 68 BSSs (6 SX Phes), 2 EHBs and 1 bright object gap object which are possible cluster members according to the proper motions from GAIA DR2.\\
    \item The comparison of FUV and NUV CMDs of the cluster with BaSTI isochrones shows that the isochrones are unable to reproduce the observed HB distribution in the UV CMDs for stars with $T_{eff} >$ 11,500 K which are known to suffer from the diffusion effects such as radiative levitation. We detect a luminosity plateau in the FUV at this temperature, suggesting the FUV CMDs to be a good proxy to detect the onset of diffusion in the BHB distribution.\\
    \item The BHB temperature distribution derived from two approaches, UVIT colour - $T_{eff}$ relation and SED fitting using Kurucz models are consistent with the spectroscopic observations. The $T_{eff}$ distribution consists of two groups with the peaks located at $\sim$ 10,300 K and 14,000 K which terminates at $\sim$ 18,000 K. The BHB stars in the hotter peak are affected by atomic diffusion. The dip between the peaks could be caused by the presence of four gaps located mainly around the G-jump.\\
    \item The temperatures and radii of the two EHB candidates and bright gap object derived from the SED fitting lie in the range 25,000 - 32,000 K and 0.15 - 0.2 R$_{\odot}$ respectively. The presence of these EHB candidates in this low density and binary rich GC could suggest binary pathways for their formation.\\ 
    \item The BSS parameters derived from the photometry and SEDs show a peak in age at $\sim$ 4 Gyr and in mass at $\sim$ 1 M$_{\odot}$.\\
    \item  The cumulative radial distribution clearly shows that the BBSSs are more centrally concentrated than the FBSSs and BHBs. The specific frequency of BSS shows a bimodal distribution suggesting that the cluster is of intermediate dynamical age. 
\end{enumerate}

\section{Acknowledgements}
We thank the anonymous referee for valuable suggestions that helped in improving the quality of the manuscript. This publication uses the data from the AstroSat mission of the Indian Space Research Organisation (ISRO), archived at the Indian Space Science Data Centre (ISSDC). UVIT project is a result of collaboration between IIA, Bengaluru, IUCAA, Pune, TIFR, Mumbai, several centres of ISRO, and CSA. This research made use of Topcat \citep{2011topcat}, Matplotlib \citep{Hunter:2007}, IPython \citep{ipy}, Scipy \citep{scipy1, scipy2} and Astropy, a community-developed core Python package for Astronomy \citep{astropy2013, astropy2018}. 

\bibliographystyle{mnras}
\bibliography{ngc288_ref.bib}
\end{document}